# Applying Ontologies and Knowledge Augmented Large Language Models to Industrial Automation: A Decision-Making Guidance for Achieving Human-Robot Collaboration in Industry 5.0

**Highlights**

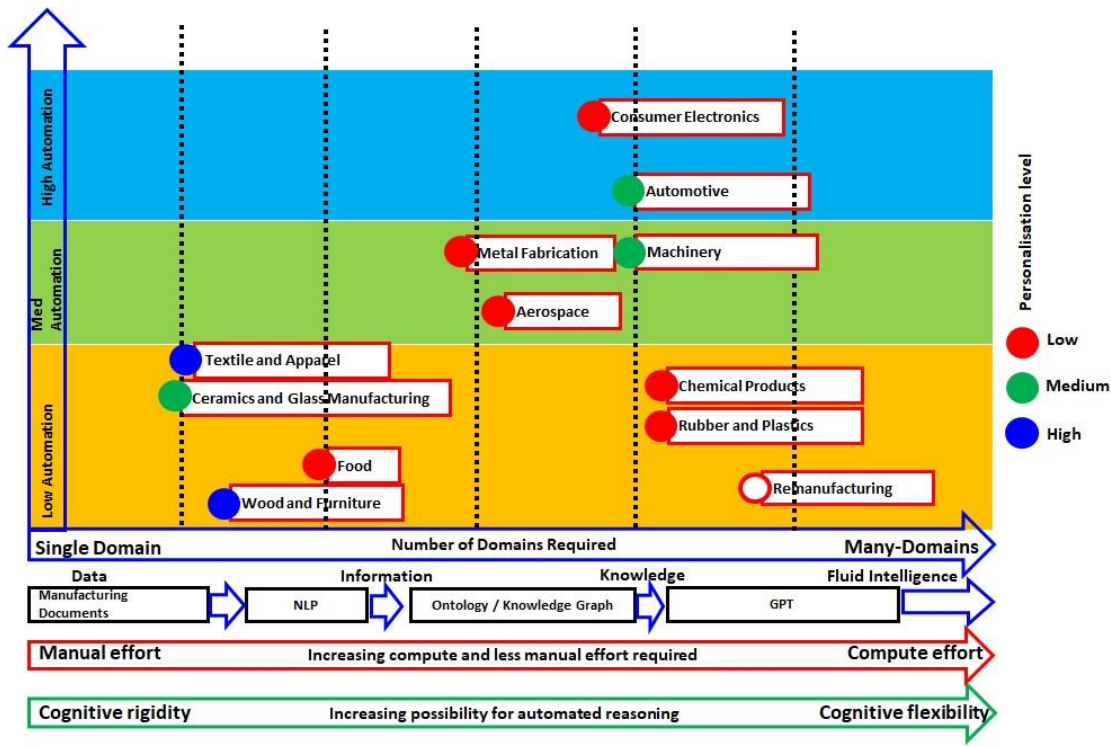

- We explore various Natural Language technologies to enhance human-robot interaction.
- We explore Language Models, Ontologies and Knowledge Graphs for various sectors.
- Various manufacturing sectors have different levels of dependency on other sectors.
- Sectors with high dependency have the most to benefit from Large Language Models.
- We provide a framework on when and which technology to choose for various contexts.

# Applying Ontologies and Knowledge Augmented Large Language Models to Industrial Automation: A Decision-Making Guidance for Achieving Human-Robot Collaboration in Industry 5.0

John Oyekan[a], Christopher Turner[b], Michael Bax[c], Erich Graf[d]

[a]Department of Computer Science, The University of York, York, YO10 5GH, United Kingdom.
[b]People-centred AI Institute, The University of Surrey, Surrey, GU2 7XH, United Kingdom.
[c]Text Inspector, Acocks Green, England, Birmingham, B27 6LG, United Kingdom.
[d]University of Southampton, University Road, Southampton, SO17 1BJ, United Kingdom.

**Abstract**- The rapid advancement of Large Language Models (LLMs) has resulted in interest in their potential applications within manufacturing systems, particularly in the context of Industry 5.0. However, determining when to implement LLMs versus other Natural Language Processing (NLP) techniques, ontologies or knowledge graphs, remains an open question. This paper offers decision-making guidance for selecting the most suitable technique in various industrial contexts, emphasizing human-robot collaboration and resilience in manufacturing. We examine the origins and unique strengths of LLMs, ontologies, and knowledge graphs, assessing their effectiveness across different industrial scenarios based on the number of domains or disciplines required to bring a product from design to manufacture. Through this comparative framework, we explore specific use cases where LLMs could enhance robotics for human-robot collaboration, while underscoring the continued relevance of ontologies and knowledge graphs in low-dependency or resource-constrained sectors. Additionally, we address the practical challenges of deploying these technologies, such as computational cost and interpretability, providing a roadmap for manufacturers to navigate the evolving landscape of Language based AI tools in Industry 5.0. Our findings offer a foundation for informed decision-making, helping industry professionals optimize the use of Language Based models for sustainable, resilient, and human-centric manufacturing. We also propose a Large Knowledge Language Model architecture that offers the potential for transparency and configuration based on complexity of task and computing resources available.

**Keywords**: Language Models, Generative Pre-Trained Transformers, Robotics, Manufacturing, Reasoning

## 1. Introduction

Historically, new technology developments have led to an evolution of manufacturing systems. Before steam power was connected to manufacturing systems, artefacts were constructed in low volumes and bespokely for customers. Craftsmen built artefacts like table or chair according to customers personal specifications and preferences. In such cases, the job was heavily manual but the quality of the product was high, bearing the skill of the craftsman. When steam power was introduced (**Industry 1.0**), it led to an increase in power being available to craftsmen to power their tools. This resulted in an increased productivity and enabled the production of more competitively priced products [1, 2, 3, 4,

5]. With the introduction of electricity (**Industry 2.0**) and the development of the automobile by Ford, manufacturing started to evolve towards standardized manufacturing of products in manufacturing cells [6]. This led to a high volume of products being produced but with low variety and flexibility in the manufacturing system [7]. Since then, the conflict or tradeoff between volume and variety has always existed in manufacturing systems as shown in Figure 1.

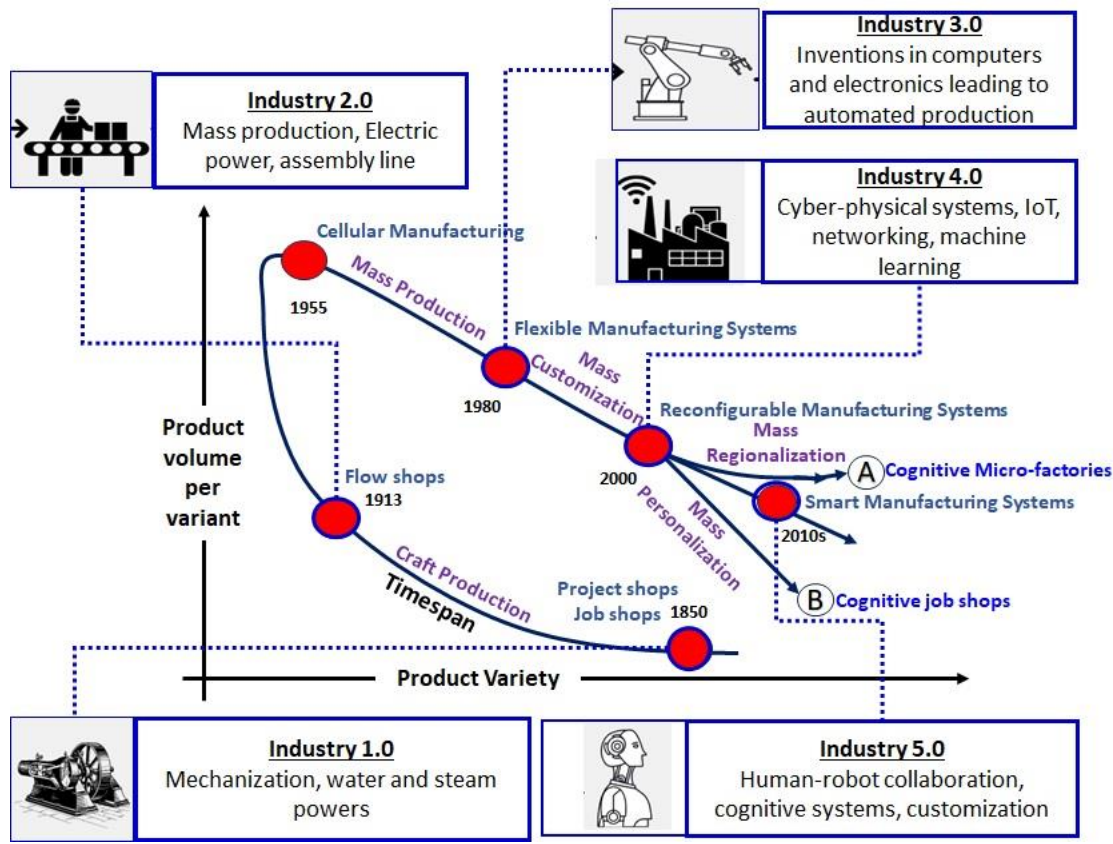

Figure 1. How technology revolutions have shaped the evolution of manufacturing systems. With the integration of Generative Algorithms with manufacturing systems, the white circles represent potential directions manufacturing systems could evolve towards. We envision Cognitive Micro-factories (A) as cottage-type factories with automation that can be flexibly and rapidly retasked for a variety of repetitive tasks. This would lead to the production of medium volume high variety products. Humans would provide oversight by ensuring that the robots have the right instructions and materials to do the task. Similarly, Cognitive job shops (B) would offer high variety as well but at lower volumes.

The development and introduction of portable computers around 1980's (**Industry 3.0**) made it possible to program a variety of instructions and hold them in a computer's memory. These instructions could then be called upon as required to manufacture a specific product thereby enabling the era of automation as well as the development of flexible manufacturing systems. Flexible manufacturing systems particularly require complex programming in order to ensure that components arrive in the correct sequence in order to manufacture a variety of products.

The internet and Radio Frequency Identification (RFID) made it possible to network manufacturing systems and obtain data from products as they passed through a manufacturing system from raw materials up to the finished product (**Industry 4.0**) [8, 9, 10]. This enabled companies to build a digital thread of various raw materials as they evolved into the final product. The amount of data generated by RFID sensors and other sensors along the supply chain, manufacturing and sale enabled manufacturers to apply Artificial Intelligence (AI) and Data Analytics algorithms to extract insights to design better customised services and products. Due to the insights offered by the collected data, it was possible to identify evolving trends in customer desires and hence reconfigure manufacturing systems in time to deal with these trends. This made it possible for manufacturers to cater more specifically to people's needs and preferences. Furthermore, such reconfigurable manufacturing systems were situated regionally in order to make them cost effective while responding faster to the bespoke needs and demand changes of consumers [11, 12].

Nevertheless, the manufacturing of bespoke goods with uniquely different components also make them difficult to recycle en-masse using a common manufacturing line. Furthermore, the use of flexible automated manufacturing techniques supported by collaborative robots with learning by demonstration techniques would not work due to the high variety in product's assembly. This makes it challenging to meet the sustainability pillar of industry 5.0.

To meet this sustainability challenge head-on, collaborative robots such as the UR5 and IWR would need to be equipped with a higher level of cognition and reasoning . This is necessary in order for them to be effective in these scenarios where flexibility is important in disassembling multi-variety and multi-component product families while conforming to industrial standards. As a result, computational technologies capable of understanding the relationships between disciplines, systems, contexts and domains will become more important in such situations as identified in the principles of **Industry 5.0** (See Figure 2) [1, 2, 14, 15]. This creates opportunities for the next evolution of manufacturing systems as shown by the white circles in Figure 1.

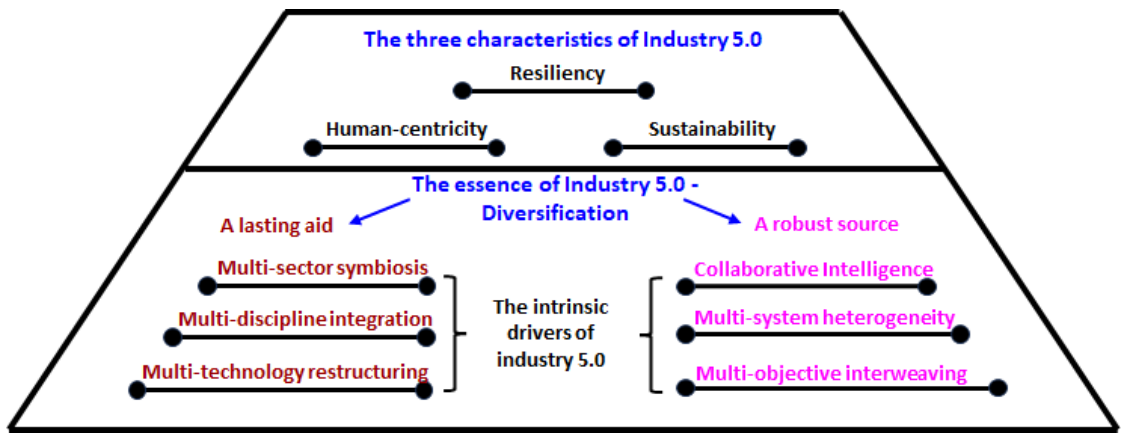

*Figure 2: The central characteristics of industry 5.0 are human-centricity in manufacturing systems, resiliency of the manufacturing system as well as sustainability in achieving productivity and the manufacturing of goods. In order to achieve these, a multi-system and*

*multi-disciplinary approach is needed to integrate various systems and disciplines together in a common collaborative intelligent architecture. This architecture would need to apply multi-disciplinary many-objectives optimisation in order to address the tradeoffs in the various represented disciplines [1].*

In driving this next evolution, advances in Artificial Intelligence (AI) in recent years has led to the development of new tools namely generative algorithms. These algorithms have been trained on massive datasets and are capable of using that information to generate new content in text, image, video and audio formats. These algorithms, have their origins in the field of natural language processing and experienced a radical paradigm shift when large-scale transformer neural network architecture were developed [16].

Large-scale transformer neural network architectures enabled the development of Large Language Models (LLM) which mined and represented huge data repositories of human generated content in an easily accessible manner. For example, ChatGPT, a chatbot based on Generative Pre-trained Transformer (GPT) technology, enables users to access a wide range of knowledge via engineered textual prompts. Now entering its 4th generation, ChatGPT promises faster responses and the possibility to mine real time and near real time data.

Human centricity is one of the main pillars of Industry 5.0 [1] and GPT technology has the potential to be incorporated within human-in-the-loop manufacturing systems towards augmenting human productivity and efficiency [17, 18, 19, 20]. For example, in addition to [21] highlighting the potential of GPT in manufacturing, [19] discussed a framework of utilizing large language models for enhancing natural language communication between humans and robots while [20] went further to investigate the use of non-verbal cues to enhance natural language during human-robot collaborations. Nevertheless, all these work propose a blanket application of LLMs in manufacturing regardless of the challenges of various manufacturing sectors and the computational resources available. Instead, in this work, we address a gap in literature of when to apply LLMs while considering that different manufacturing sectors have different challenges and hence have different requirements.

We propose a methodological framework to guide practitioners and decision makers in knowing **when to choose LLMs from other available natural language technologies (such as natural language processing techniques, Ontologies and Knowledge Graphs)** and under what conditions to apply them while achieving their purposes **at lower computational costs, greater transparency and with reduction in data privacy fears**.

Towards this, we present an overview of the origins of Generative Pretrained Transformers (GPT) in natural language processing (NLP) techniques and how NLP has been used in supporting Human-Robot Collaboration in manufacturing. We then apply lessons learnt in the applications of NLP in robotic manufacturing applications to support potential human-robot centric GPT applications in manufacturing. We also present a framework to support decisions on applying GPT in robotic manufacturing systems. Furthermore, we discuss potential opportunities and challenges that exist in doing so. More specifically, we aim to address the following Research Questions:

- **RQ1:** What form might an information processing architecture (cognitive architecture) take in order to achieve human robot collaboration especially in the context of the human centricity characteristic of industry 5.0? It is important to understand how the computational models would interact with planning and control in such architectures to achieve the goals of industry 5.0 including collaborative intelligence and multi-sector symbiosis (See Figure 2).

- **RQ2:** What level of computational model (including language models) should be used for robotics or automation in various manufacturing use cases? This is especially true as large language models demand energy intensive computations during training as well as during fine-tuning and not all manufacturers have access to these type of resources. Furthermore, large language models have a long inference time when compared with Natural Language Processing computational models. Computational models based on Natural Language Processing are computationally cheaper providing benefits as well.

- **RQ3:** How do we determine which manufacturing task stands to benefit the most from the application of Large Language Models? There is a tradeoff between the complexity of the task and the computational model needed to solve it. Knowing that tradeoff will support decisions during the building of computational models for human-robot collaborative systems.

- **RQ4:** Since resilience and sustainability are key pillars in industry 5.0, what opportunities are there for various manufacturing sectors when applying automation including for remanufacturing?

Furthermore, these Industry 5.0 automation technologies would need to be capable of explanability and human action understanding in other to achieve human-centricity. This would enable humans to make use of their inherent adaptability and flexibility in such **Industry 5.0** inspired manufacturing systems [17]. In these manufacturing systems, creativity of humans, simultaneous co-sharing of work between humans and robots as well as sustainability are central themes. As a result, natural human communication techniques with automation would become important in these new manufacturing systems. In the next few sections, we explore how the field of natural language processing for automation has progressed and applied towards achieving this goal.

## 2. Natural Language Processing in Manufacturing Robotics and Automation

Natural communication between humans and robots has the potential to increase trust levels in human-robot systems [22, 19, 20]. However due to the disparity between humans (a self-sustaining meta-cognitive biological system) and robots (a non-sustaining mechanical system), collaboration between these entities on tasks has been difficult to achieve. Robots need atomic commands (such as move the motor by 3rad/sec) that are suited to their mechanical systems. This is different to the way humans communicate (move to the table) [23]. As a result, researchers have been investigating how to fluidly decode human natural language into atomic commands for robots through the use of cognitive architectures (information processing architectures) which mimic the way humans process information from the environment [24, 25, 26].

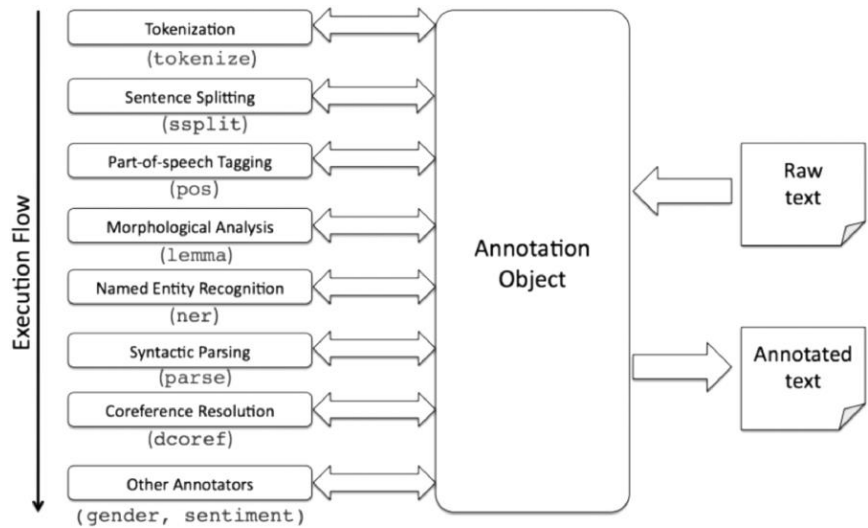

*Figure 3. Stanford Natural Language pipeline for extracting annotated text [23]. The input to this framework is raw text which is decoded into its constituents (via tokenization, part of speech tagging, entity recognition etc) and outputted as annotated text for further computational processes upstream.*

## *2.1 Decoding Human Natural Language for Robots*

In [27], Manning et al. developed the Stanford Natural Language Processing (SNLP) architecture (See Figure 3) to convert natural spoken speech into sentences and annotated text. The annotated text was output into a Resource Description Framework (RDF) language which is a general framework for representing interconnected data on the web and enables the standardization of data exchange based on the relationships between objects and items [28].

In [29], Recupero et al. used the annotated text output of SNLP to construct actions for a Nao robot, a small humanoid robot. SNLP enabled parts of speech tags such as Verbs, Adverbs, Adjectives etc to be extracted from user spoken sentences. It was also possible to extract the syntactic dependency relationships, that is Subject-Verb-Object, between objects in the spoken sentences. The objects (e.g arm) in the sentences were linked to adjectives (e.g left or right) and then finally linked to the Verb token which defined what to do with the object. In their work, a typical command could be: "Nao, move up your left hand" with the Subject being Nao; move (verb), up (adverb), left (adjective), hand (object). This approach makes it possible to mine existing human readable resources for recipes and instructions to achieve various end goals and enables deterministic actions to be carried out by a robot in parallel with a human [30].

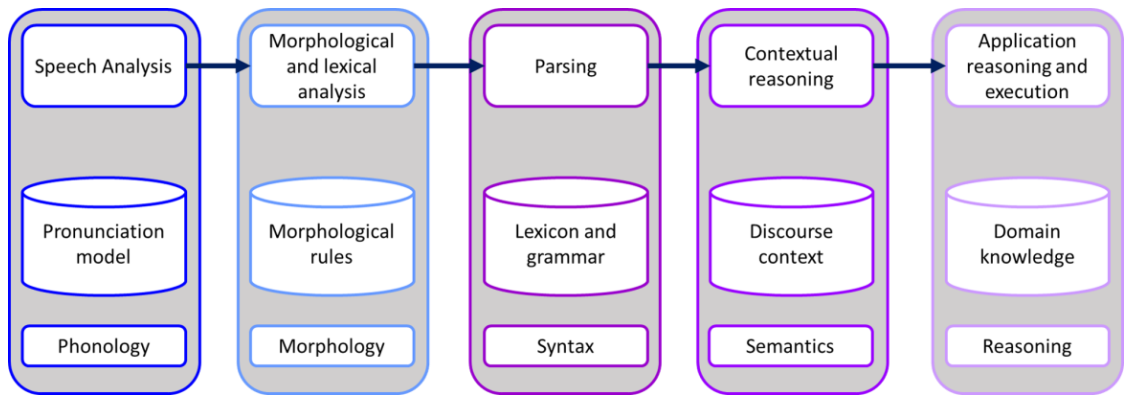

*Figure 4. A knowledge of the domain is essential in ensuring effective reasoning by the AI agent.*

In these approaches, the robot is able to translate the human written instructions into continuous action sequences in order to complete tasks. In [31], Markelius et al. created an hierarchical system to achieve this. Their hierarchical system was made up of three layers: task layer (top layer), semantic motion layer (middle layer) and motion primitive layer (bottom layer). By combining written instructions with visual cues from the environment, they were able to use the current state of objects in the environment in the semantic motion layer to generate motion language. The motion language consisted of the visually detected object structure and its six-dimensional state (position and posture). This was then passed onto the motion primitive layer to generate the required actions in the form of robot motion primitives [31].

## 2.2 Applying contextual NLP in industrial robotics

However, the context of the current task as well as its domain is important as can be seen in Figure 4. In [30], a language-based planning framework enabled robots to solve sequential manipulation tasks that require long-horizon reasoning. This enabled the possibility of capturing the context of a current task and then for corresponding actions to be orchestrated accordingly.

In [32], Ho et al. investigated the use of cognitive linguistic constructs such as spatial relationship constructs to develop a Natural Language Unit for converting complex natural language instructions into actions to be taken by a robot. The use of language grounding in which the mapping of natural language instructions to robot behaviour was studied in [33]. They discussed that such systems should be capable of ensuring the right selection of behaviour as well as contextual generalisation beyond the existing commands and configurations seen at training time. This resulted in a proposed language grounding hierarchical planner for robot control that can perform language abstraction. Through the utilization of Single-RNN and I-DRAGGN models, they were able to ground and execute spoken natural language instructions in real time while responding to a wide range of natural language commands. Another approach investigated by [34] made use of semantic compositionality to achieve contextual generalisation. Semantic compositionality is a developmental robotics model that models how infants slowly acquire language in various stages. **However, achieving contextual understanding within and across diverse**

**domains is quite challenging with NLP due to amount of manual effort required.** This problem becomes more pronounced as the number of domains increase. As a result, a different approach is needed.

## 3. Ontology and Knowledge Graphs in Manufacturing

The development and application of Ontologies and Knowledge Graphs in manufacturing were partly motivated by the technical challenges of using NLP techniques to achieve contextual understanding within and across manufacturing domains. These tools build on NLP techniques for tokenization, entity relationship extraction and so on during the early phases of their development pipelines. Knowledge Graphs represent real-world facts and the semantic relationships between them in the form of triplets: (subject, predicate, object) OR (head, relation, tail) (see Figure 5 [35]). This makes it a structured way to organize and represent knowledge, typically in the form of nodes (representing entities or facts) and edges (semantic relationships). In addition, a knowledge graph can also embed the values or attributes that the nodes can take. As a result of its inherent structure, a knowledge graph offers the opportunity to make querying and analyzing of domain data and information easier.

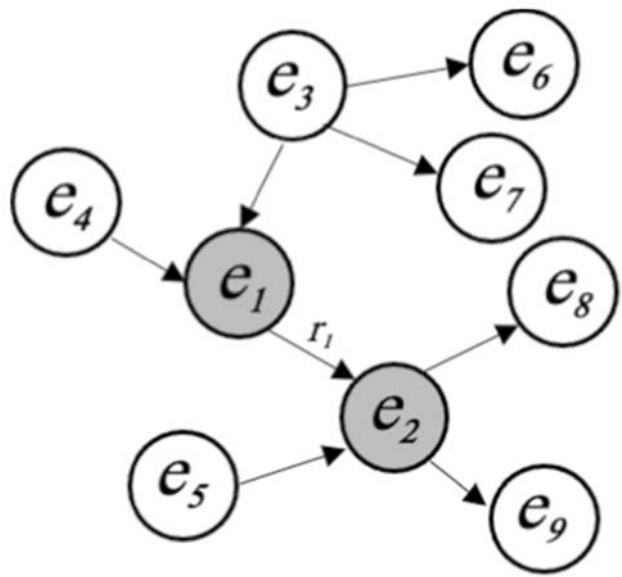

*Figure 5. A representation of a Knowledge Graph in which $e_1$ and $e_2$ are connected by the relationship $r_1$. This results in a triplet ($e_1$, $r_1$, $e_2$).*

On the other hand, an ontology is a formal representation of knowledge that defines the concepts and relationships within a specific domain. It typically includes a taxonomy of classes, properties, and instances, along with the constraints and rules governing them. As a result, ontology construction is often required in order to achieve a knowledge graph. In the sections below, we will use the terms ontology and knowledge graphs interchangeably.

## 3.1 Constructing Ontologies in Manufacturing

The key stages required in the construction of knowledge graphs, from the initial stages of its development all the way to its deployment in AI Systems and other application fields is shown in Figure 6. Through Parts Of Speech (POS) tagging, Natural Language Processing offers the potential to convert human understandable text from manufacturing documents into information and knowledge that can supplemented with data from sensors on the manufacturing floor. In the context of Figure 6, entities and the relationships between them can be extracted using POS ① thereby contributing to the Knowledge Acquisition step. This results in an automatically annotated text with semantic tags. However, this step might not completely extract all the information required and is also dependent on the type and number of manufacturing documents available.

The knowledge Graph completion step ② improves the quality of the Knowledge Graph through a link and entity prediction step which is then validated by the user using external sources of information. In the manufacturing sector, there are many siloed domain-specific ontologies such as the one provided by the OPC Unified Architecture (OPC Foundation, 2017); digital twins metamodels for Digital Twins (Plattform Industrie 4.0, 2020); production ontologies based on BFO (Industrial Ontologies Foundry, 2020) to mention a few [36].

In order to enrich ontologies from such sources, the knowledge fusion step ③ further uses information from other domains to augment and enrich the generated knowledge graph. In [11], Mo et al. proposed a framework that combined translations, domain-specific vocabularies and inconsistency checks with syntactic terminological as well as structural analyses. This enables integration of knowledge representations from various domains that have been formalized in the Web Ontology Language. The approach of developing an heterogeneous domain ontology ensures that the reasoning produced by agents is much richer than single domain ontologies. This makes for a much rounded artificial agent. **Nevertheless, this would also increase the complexity of creating the heterogeneous ontology as well as its size on the disk. Furthermore, there are still research challenges in merging various ontologies into a single representation** [11].

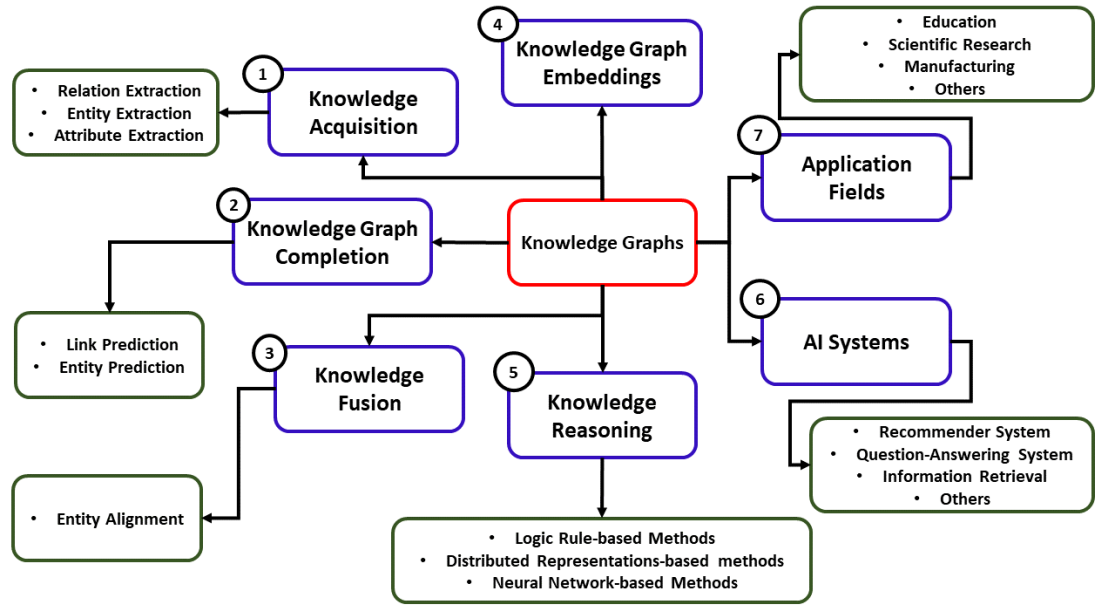

*Figure 6. Concepts in Knowledge Graphs and key stages from development to deployment [35].*

Depending on the need, techniques such as Long-Short Term Neural Networks, Graph Neural Networks or other similar methods④ can be used to create learnt models that embed knowledge into a low dimensional space. Other techniques can include the development of logic rules that embed the knowledge of the domain [37]. In the knowledge reasoning⑤ step, models and rules are interrogated by using queries or data obtained in real time either in the form of text or numerical data. At this stage, the user validates the results returned by Knowledge Graph before deployment into AI systems⑥ in various applications⑦.

In such a setup, Artificial Intelligence agents carry the heavy load of computing large sources of data from multiple modal sources. The derived knowledge and insights can then be visually presented to humans in the form of ontologies and knowledge graphs. The presentation in an ontology or knowledge graph enables humans to make inferences, query the AI agents for more information and make use of their visual cortex's brain capacity to easily extract information from the knowledge graph [11]. This is used to improve the efficiency and productivity of workers as well as manufacturing lines in various factories. Once an ontology is created, this can be used for reasoning to produce previously unknown information regarding the domain based on the relationships between the entities in the domain. Developed knowledge graphs are stored as Resource Description Framework (RDF), graph database-based storage or Ontology Web Language (OWL) which is often used for Ontologies. The use of graph databases focus on efficient graph queries and search as well as use attribute graphs as the basic representation. In this case, entities and their relationships contain attributes making it easier to reason about the domain.

Nevertheless, it is not only the knowledge graph of the domain or related domains that is required for a robot to make use of knowledge graphs in their reasoning. knowledge graphs are also required of the robot's morphology, skills and its capabilities itself. This is important in manufacturing where different types of robots and automation are used for different tasks.

### *3.2 Applying Ontology in Manufacturing Robotics*

Developing multi-domain ontology-based architectures enables AI agents such as robots to reason about concepts related to their domain tasks [11, 38]. In [29], an action command ontology for the humanoid robot Nao was developed. This contained all the physical movements that the humanoid robot can perform, its body parts as well as their related words and synonyms. The advantage of this approach is that it supports communication variability and preferences in human subjects.

In [39], Hu et al. developed and made use of a product's end of life information along with a rule-based reasoning method, to develop a human-robot collaboration disassembly ontology. Their ontology automatically generates the optimal disassembly sequence and scheme while combining supportive rules with the ontology model. These rules are customised disassembly-related rules which ensure that the constraints related to the product are adhered to during the disassembly task of EoL products. However, in their architecture, it was not clear how they took into account the knowledge of the robot's morphology into consideration in their ontological mapping.

This was addressed in [40] where Zheng et al. developed a knowledge-based approach to automatic program generation for robotic manufacturing systems. In their work, they fused ontological knowledge models from robotic manufacturing, basic instruction units for robotic programs and workpiece product models to achieve a semantic description of the automated manufacturing system. They then built a rule-based reasoning mechanism to infer relationships between the basic units of program required to do various task on a product's workpiece. This was then used to automate the generation of a program relevant to the task at hand while dealing with various product types.

David et al. also proposed a collaborative agents manufacturing ontology (CAMO) framework in which humans and robots are seen as a self-organising team network [37]. Building on the Web Ontology Language (OWL) they incorporated models of agents, manufacturing systems, and various interaction contexts into their knowledge base. This enabled the opportunity to generalise to different assemblies and agent capabilities. Furthermore, they used semantic representations of agent capabilities to achieve a dynamic consensus-driven collaboration between humans and robots. These examples show that Ontologies and Knowledge Graphs are still relevant and powerful on the manufacturing landscape. Furthermore, the use of knowledge fusion enables designers to pull together knowledge from various domains specific and relevant to a particular manufacturing application. **Nevertheless, for Knowledge Graphs and Ontologies, developing a methodology that can efficiently combine multiple to many domains into a single Knowledge Graph or Ontology for artificial reasoning still needs research**. This is because the amount of manpower required to develop an Ontology grows as the number of domains to integrate increases. In this case, Large Language Models and

Generative Pre-trained transformers provide a coherent way of addressing the challenge of integrating knowledge from many domains.

## 4. Large Language Models and Generative Pre-trained transformers

Understanding the history of Language Models is important in order to see how they have evolved over the decades (Figure 7), what categories of Language Models there are, the compute resources required to run them and where they might be useful in human-robot collaborative manufacturing systems.

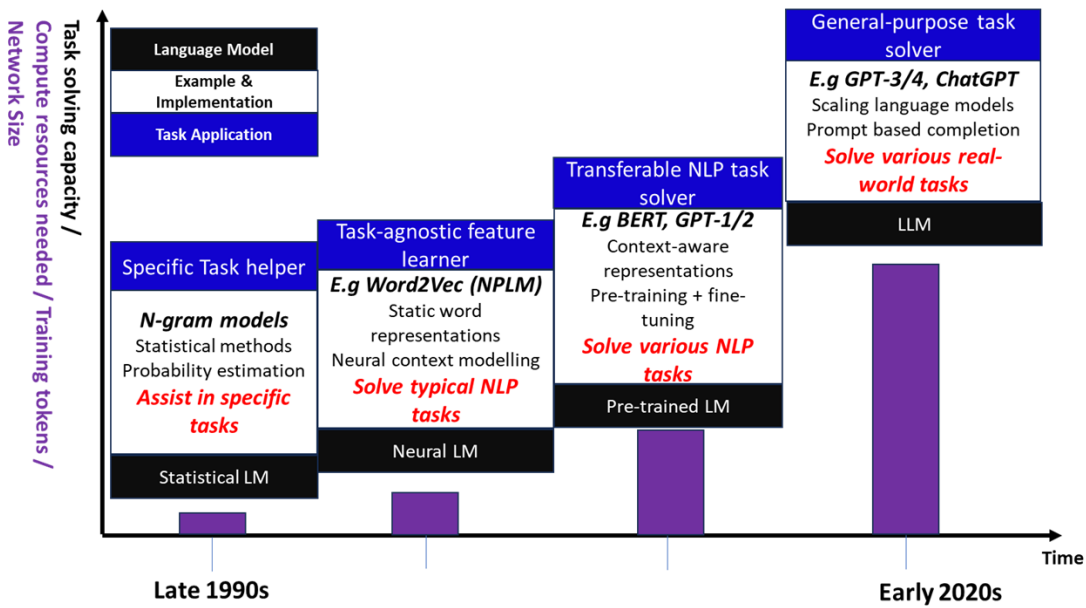

*Figure 7. Evolution of Language Models starting from applying NLP with statistical modelling up to LLMs. LLMs enabled more global context-aware processing compared to local prediction of the next word sequence in Statistical Language Models. However, the computational resources required to train the models increased exponentially.*

### *4.1 A brief history of Language Models*

Language models aim to use probabilistic approaches to predict and hence generate the next sequence of words or tokens from a previous set of words. The length of the previous set of words provided context and resulted in *n*-gram language models. Due to the nature of this task, a combination of Markov and statistical approaches were used. This resulted in **Statistical Language Models** which were successfully but limited by their inherent Markov property and the n-length (or n-gram) they used for inference [41, 42]. As the domain or task became more complex or multi-dimensional, the number of potential next states increased exponentially. In order to solve this problem, universal approximators such as neural networks were used. The resulting **Neural Language Models** [43, 44] made use of various variants of networks such as long-short term neural networks (LSTM) and recurrent neural networks (RNNs) which were capable of using their recurrent nature to influence the next predicted state or set of words [45, 46]. The nature of Long-Short term neural networks also enabled context to be inferred using local (Short) and long (semi-global) memories thereby enabling the Language Model to keep track of the short term sequence of words while being influenced by the overall context of the user-defined task.

The development of Bi-directional LSTMs (BiLSTMs), which consisted of two LSTM layers ; one for processing input in the forward direction, the other for processing in the backward direction and the development of self-attention mechanisms ushered in the Transformer Architecture. This was a larger Neural Network compared to former Neural Networks. Architectures such as the Bidirectional Encoder Representations from Transformers (BERT) resulted in the development of **Pre-trained Language Models** that could be fined-tuned for specific tasks. This set the scene for Generative Pre-trained Transformers (GPTs) that were even larger Language Models driven by the goal of achieving Artificial Generalised Intelligence [49, 50].

**Large Language Models** are recent additions to the Language Modelling and Natural Language Processing scene [51, 52]. They are trained on massive text data using unsupervised learning objectives and are large scale neural networks with the number of parameters ranging into the hundreds of billions compared to architectures such as BERT that had parameters of around 330 Million [49, 53, 54]. In [55], Radford et al. showed how a Transformer Architecture could be trained on a huge amount of textual data towards **predicting and generating** the next word in the context of a sentence as well as understanding the structure and patterns of natural languages. In [53], Zhao et al. discussed how LLMs are made up of smaller language models (SLMs), similar Transformer architectures as well as pre-training objectives. They discussed how the emergent property of LLMs and their capability increases as the size of models (or number of small language models), data and total training compute increase. By emergent, they meant the abilities that arise in large models but are not present in small models. This emergent ability occurs after scaling up of a Transformer Architecture crosses a certain value leading to the generation of less random results by small networks [56].

The emergent property enables LLMs to solve complex tasks using self attention mechanisms and multiple reasoning steps. It also enables them to generalise and perform well on previously unseen tasks. As a result, through fine-tuning, a pre-trained transformer architecture can be deployed to various domains. **This ability equips the pre-trained models to share generalised knowledge across domains with domain specificity left to the user.**

## *4.2 Applying Large Language Models in Manufacturing Robotics*

An advantage of Generative Pre-trained Transformers or Large Language Models (LLMs) is that they lower the barrier for less technical people to interact with Artificial Intelligent Natural Language Processing Units while using them to carry out sophisticated tasks. In [30, 34], Lin et al. and Singh et al. proposed the ProgPrompt and Text2Motion architectures respectively. These architectures apply LLM's strength in world knowledge and programming language understanding to generate text based robot executable plans. This is a new approach in industrial robotics. For example, in [57], Li et al. fine tuned the state of the art pre-trained Bidirectional Encoder Representations from Transformers (BERT), to understand the desired intents of a human operator and to convert the understanding into robot commands accordingly.

**Nevertheless, there are still challenges in converting the output of these LLM architectures into domain-specific actionable and interpretable commands for robots or autonomous systems.** Furthermore, prompt engineering, to elicit the right executable plan or action sequences for the robot, is currently challenging [58, 59]. **This is particularly true in manufacturing systems where product manufacturing flexibility and variation is paramount.** In such cases, the planning component of the robot must be able to perceive the current state of events in the workspace and constrain its actions during task and motion planning (TAMP)(See Figure 8) while understanding the current domain context within which it is operating.

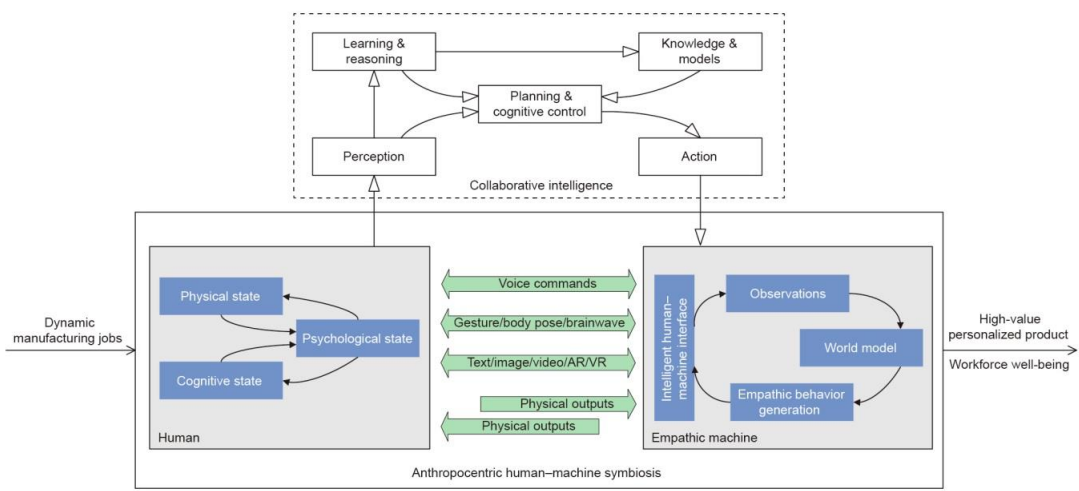

*Figure 8. An industry 5.0 inspired anthropocentric human–machine symbiosis framework for supporting ultra-flexible manufacturing systems . In this framework, the physical, cognitive and psychological state of the human needs to be digitized and decoded in order for a robot or machine to understand and serve human needs in a human-centric manufacturing system. Such a framework would serve a human in a job-shop environment in which highly dynamic manufacturing jobs would need to be completed to achieve high-value personalised products.*

In [61], Yoshikawa et al. applied LLMs to support the generation of autonomous chemistry experiments. Their architecture as shown in Figure 9 and Figure 10 was able to convert Natural Language prompts into executable robot instructions. They combined a domain specific description language with the generated instructions from a LLM in order to validate as well as constraint the output of the LLM into chemistry specific instructions. This approach ensured syntactically valid programs in a data-scarce domain-specific language that incorporates environmental constraints. In order to be able to generate this domain-specific language or constraints, it would need to be informed by the ontology of the domain.

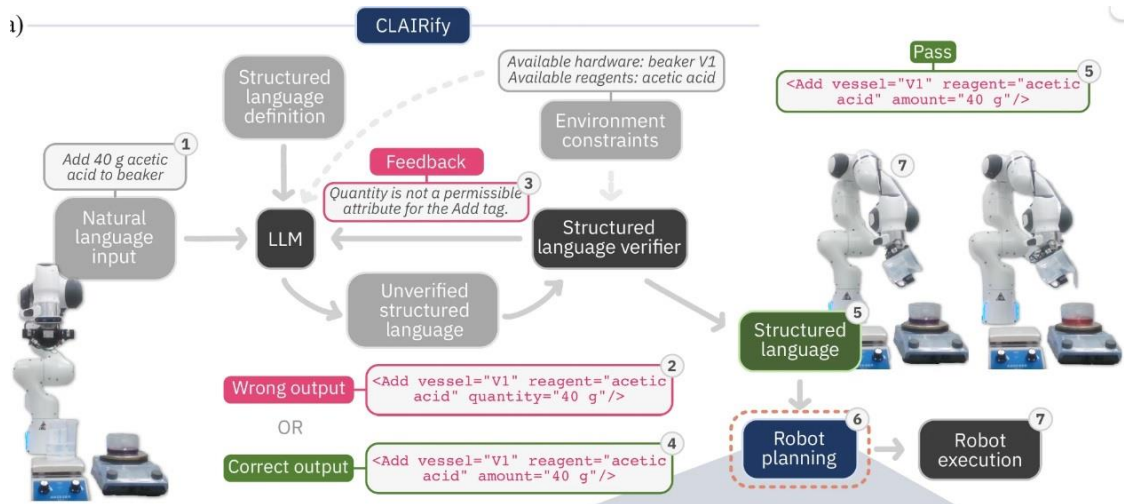

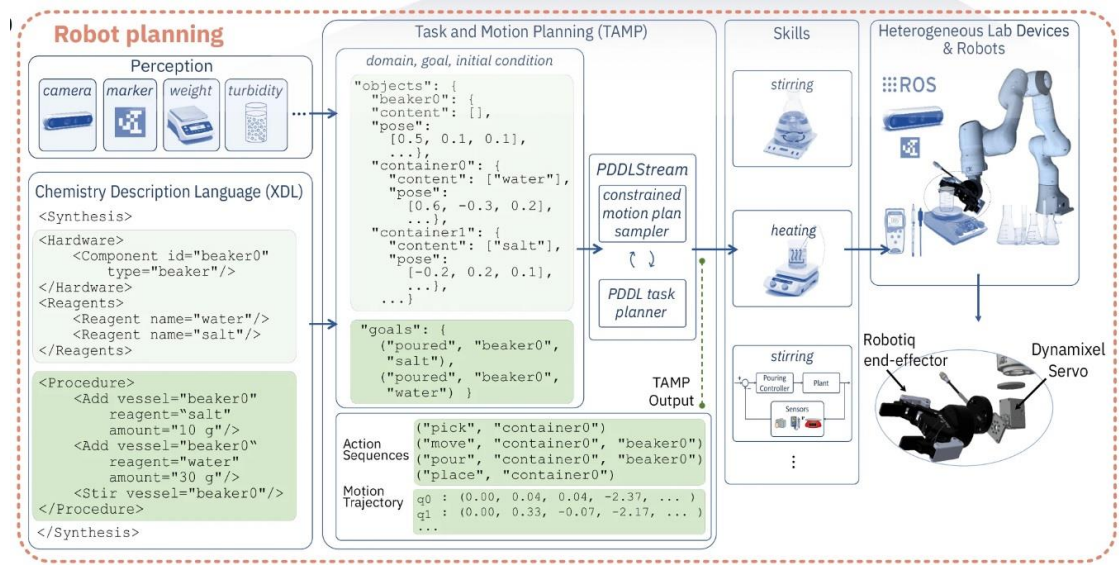

Furthermore, the results generated by LLMs are plagued by the black box effect in the sense that no one knows how the results were generated and if they can be trusted. This is especially true as LLMs are known to hallucinate and generate wrong information [53]. Towards solving this challenge, the work in [61] is unique in the sense that it uses domain-specific description language to support humans in understanding what the robot is doing. In [62], Holland et al. made use of LLMs to develop an automated fiber process planning system for composite structures. Similar to , their approach was augmented with process planning knowledge thereby enabling it to handle variations in cycle time and resource allocations.

The use of domain-specific description language opens up the possibility of increasing trust levels in LLM generated results. Another approach that is being investigated in literature is the use of human understandable linear and decision tree models to augment the outputs of LLMs. This has been used in [63] for text classification but not yet in human-robot collaboration scenarios.

As a result of the above findings, we propose that the learning & reasoning module in Figure 8 would require a domain or context specific description language in order to increase transparency and trust levels. Such an architecture would better support human robot collaboration especially in the context of the human centricity characteristic of industry 5.0 thereby addressing **RQ1**.

## 5. Experimental Evaluation of Computational Models

In order to further address **RQ1** and the rest of the research questions, we conducted an experimental evaluation of the computational models relying on Natural Language Processing, Knowledge Graph and Large Language Models.

### *5.1 Evaluation Metric*

We developed evaluation metrics based on the TOEFL (Test of English as a Foreign Language) Integrated Writing Task. This test is used to assess students' ability to read and then combine information from different sources into a coherent whole using writing. Taking inspiration from this test, we applied the following metrics in order to assess the suitability of computational models across the domains of Textiles, Electronics and Remanufacturing (these domains were chosen because of the differences in the level of dependency on other sectors- See Figure 17):

- **Content Accuracy, Semantic and Syntactic Relevance:** This assessed a model's ability to capture the key points from reference texts in the domains as well as its accuracy in capturing the relationship between the various domains represented. In assessing this, we made use of Global Vectors for Word Representation embeddings (GloVe) [64] to measure the semantic and syntactic relevance between text generated by a model and the reference text. This measure is useful in assessing if the text generated by the model captures the essence of the domain we are interested in.

- **Coherence and Organization:** This assessed the model's ability to provide a clear logical structure and connection between ideas as well as the effective use of transitions between contexts and domains. In measuring transitions, the model's output was checked for connection words like "Furthermore", "however", "therefore", "moreover", "in addition" and "furthermore". We also used cosine similarity to measure the cosine similarity between the text generated by a model and a reference text.

- **Instructional Use Score:** We introduce this metric in order to assess how well a human can understand the text produced by a model for use on a particular task.

- **Inference speed and model size:** For each of the computational models, we measured the inference time between the prompt being given to it and the results being generated. This is useful in order to measure the model's responsiveness and how that would impact robots decision making in real time.

The reference text was created by constructing a corpus for each domain. The corpus was assembled by using sciencedirect to compile a list of over 80 review and research papers for each domain. We used the search terms $[domain] + [manufacturing]$ to identify papers. We tend extracted the text from the papers and preprocessed it by removing academic references, filler words (like introduction, abstract etc), author names, country and phone numbers.

## 5.2 Experimental Setup

In evaluating the computational models, we created a virtual machine running Ubuntu 24.04 with a RAM of 16GB, 2 core-processors, no Graphical Processing Units and a hard disk of 70GB. We tested the computational models' outputs across the domains of Textiles, Electronics and Remanufacturing by making use of the following task prompts respectively:

- Construct a cotton T-shirt, starting from fiber production to fabric to finished garment.
- Assembly a mainframe from electronic disks and keyboards.
- Convert an internal combustion engine vehicle into an electric vehicle.

These prompts were developed with a view to generate instructions that a collaborative robot could follow to complete a task. They were fed through three computational models: Natural Language Processing (NLP) computational model (See Figure 3), WordNet (to represent a Knowledge Graph) and Generative Pretrained models (GPT2, GPT2-medium, GPT2-Large, GPT2-XL). The model sizes are shown in Table 1.

When prompting the **NLP model**, we used keywords (Nouns and Verbs) to search the reference text and return found sentences. In developing a Knowledge Graph, we made use of WordNet in the NLTK library. WordNet is a generic Knowledge Graph that captures the semantic relationships and definitions between various objects [65].

For the WordNet computational model, we made use of the entire prompt after extracting the parts of speech of the prompt and applying sense disambiguation. Each word in the prompt was replaced by its WordNet's definition over two iteration thereby producing a series of text and sentences that contained results of the prompt.

Similar to [20], the Generative Pretrained (GPT) Models were used out of the box in order to provide a fair testing ground for all of the computational models used. We made use of GPT models in this work because they make use of decoder only architectures for generating text. Other transformer architectures such as BERT models use only encoders, while transformers like T5 (Text-To-Text Transfer Transformer) and BART (Bidirectional and Auto-Regressive Transformers) use encoder-decoder architectures for summarisation and translation tasks.

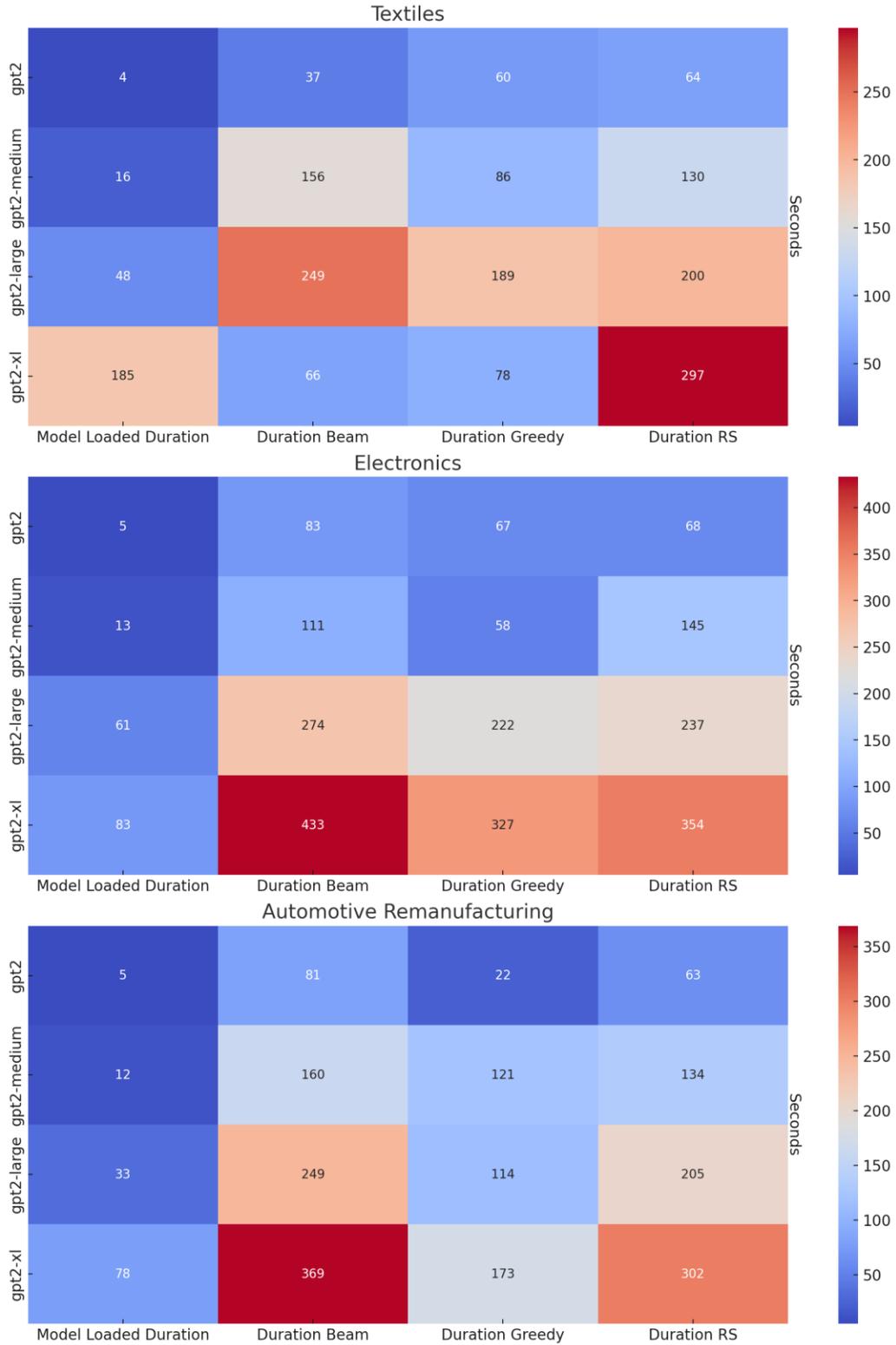

Figure 11: Time taken by various GPT models to load and to make an inference.

## 5.3 Evaluation Results

In this section, we assessed the outputs of the computational models according to the metrics we discussed earlier.

**Inference time and model size:** Figure 11 shows the time taken for the GPT computational models to make an inference with values ranging from 37 seconds in GPT2 up to 433 seconds (7 minutes) in GPT2-XL. Across each of the GPT2 models, we applied three inference search strategies: Beam, Greedy and Random Sampling. In generally, the greater the model size, the longer the inference time. The WordNet model inference time had a maximum value of 10 seconds while that of the Natural Language Processing model depended on the size of the reference text it needed to go through. This could be made faster through the use of parallel processing computational techniques in the future.

| Model | gpt2 | gpt2-med | gpt2-lge | gpt2-xl | WrdNet | NLP |
|---|---|---|---|---|---|---|
| Size | 0.55G | 1.52G | 3.35G | 6.43G | ~0.012G | >0.012G |

*Table 1. Model sizes for different architectures. G stands for Gigabytes; Med- Medium; lge- Large and xl- eXtra Large*

Furthermore, the larger the size of the model, the longer it took to load into RAM as shown in the first columns of Figure 11.

**Content Accuracy, Semantic and Syntactic Relevance:** Figures 12 to 14 show the results of the models in generating instructional text that fit the reference text of the domain they are currently prompted with. The glove embeddings metric show that all the models scored highly for semantic and syntactic relevance with WordNet and NLP comparable with the GPT models across all domains and in some cases scoring highly than them.

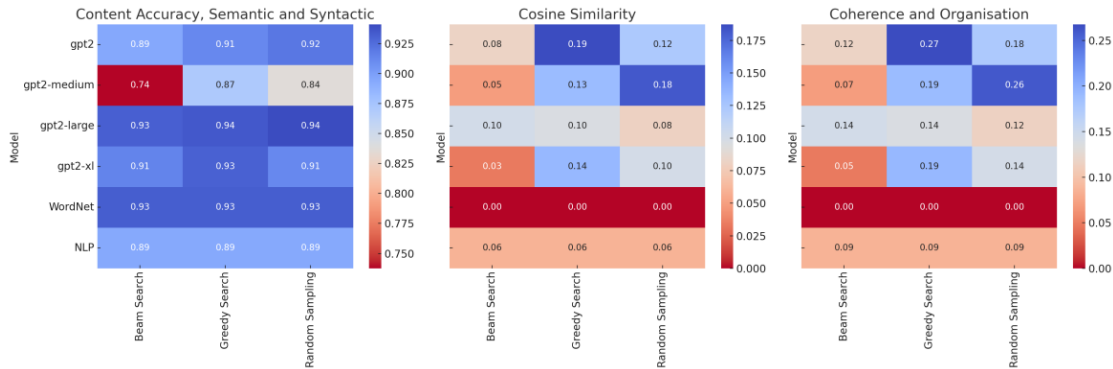

*Figure 12: Performance of the computational models in the Textiles domain. It should be noted that for comparison purposes, we replicated the values of the WordNet and NLP models across Beam, Greedy and Random Sampling of the GPT models.*

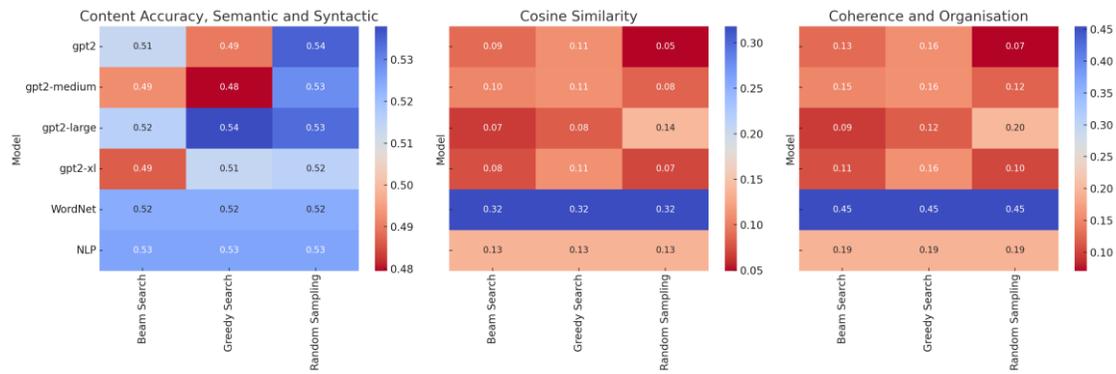

*Figure 13: Performance of the computational models in the Electronics domain. It should be noted that for comparison purposes, we replicated the values of the WordNet and NLP models across Beam, Greedy and Random Sampling of the GPT models.*

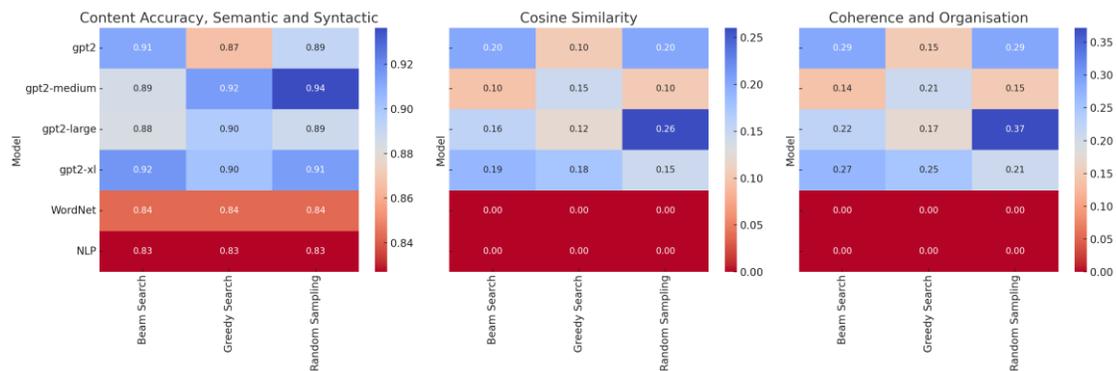

*Figure 14: Performance of the computational models in the Remanufacturing domain. It should be noted that for comparison purposes, we replicated the values of the WordNet and NLP models across Beam, Greedy and Random Sampling of the GPT models.*

**Coherence and Organization:** In terms of coherence and organization, NLP computational model produced better results than WordNet as well as GPT2-XL's beam search and GPT2-medium's beam search strategy in the **Textiles** domain. In the **Electronics** domain, the WordNet model performed the best across all models with the NLP model taking second place. However, when using the computational models for a **Remanufacturing** task, the GPT models performed better with GPT2-Large's Random Sampling scoring the highest value. Furthermore, the scores of the GPT models improved overall in the **Remanufacturing** domain task compared with **Textiles** and **Electronics** domains.

**Instructional Use Score:** As mentioned earlier, we introduced this metric in order to assess how well a human can understand the text produced by a model for use on a particular task. In the future, the generated text will be converted into instructions for robots to make use of. We assigned scores of 0 to 1 to the text generated as shown in the Appendix A to C. Figure 15 gives an overview of the performance of the generated text across the sectors. For the ***Electronics*** domain, the WordNet model's performance scored highly while in the ***Remanufacturing*** domain, WordNet results were comparable with some of the GPT2 search strategies. Nevertheless, the GPT2 models performed better than the WordNet model when using Greedy search strategies. In the ***Textiles*** domain, the NLP model performed better than the WordNet model as well as better than the GPT2 models over half of the time. For the ***Remanufacturing*** domain, the GPT models also generated text that covered legislation needed to support the conversion of a combustion engine vehicle into electric without being explicitly prompted.

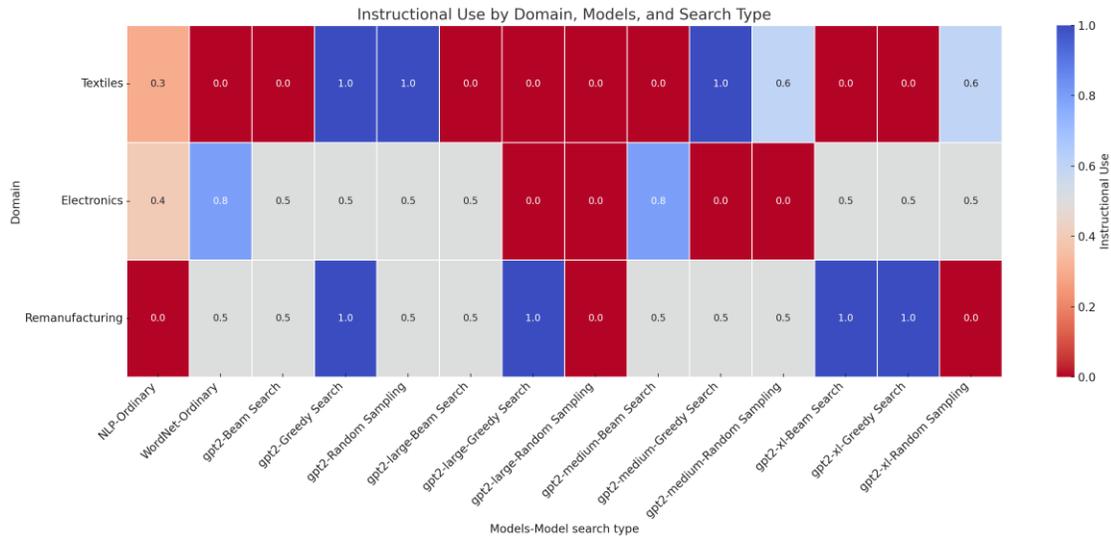

Figure *15: Assigned scored based on how well the generated text is understandable and can be used by a human as instructions for the task.*

## 6. Discussion

### *6.1 Which computational model to apply: NLP, Ontologies, Knowledge Graphs or LLMs (RQ2)?*

NLP, Knowledge Graphs and Ontologies are well established [66, 67] and have been used in the research community for decades. Furthermore, as seen in the references of [39, 40, 68], they are still highly relevant in the era of LLMs and offer the advantages of transparency in the understanding of their inner workings as well as explanability thereby increasing trust in these systems. By their nature, Ontologies and Knowledge Graph architectures can be used in the development of ethical and trusted human-robot collaborative systems. Through knowledge fusion techniques, knowledge from manufacturing, ethical, legislative standards, health and safety as well as other relevant domains can be incorporated into an Ontology or Knowledge Graph iteratively thereby supporting the understanding and debugging of developed human-robot collaborative systems. Such heterogeneous Knowledge Graphs or Ontologies could approach the ability of Pre-trained Large Models without the compute requirements to train hundreds of billions of neural weights.

| Name of Model | Model size | Compute resource required | Manual effort required | Contextual reasoning | Transparency | No of domains captured | Reasoning and generalisation ability |
|---|---|---|---|---|---|---|---|
| Ordinary NLP | 🔵 | 🔵 | 🟢 | 🔴 | 🔵 | 🔴 | 🔴 |
| Ontologies & Knowledge Graphs | 🟡 | 🟢 | 🔵 | 🟢 | 🔵 | 🟢 | 🟡 |
| Large Language Model | 🔴 | 🔴 | 🔵 | 🔵 | 🔴 | 🔵 | 🔵 |

🔵 🟢 🟡 🔴

More favourable ──────────────► Less favourable

*Figure 16: A comparison of models that can be used in the knowledge and models module of the human-robot collaborative intelligence in Figure 8. The colors are in ascending order of Blue, Green, Orange and Red with Blue meaning favourable and Red unfavourable.*

However, their construction would require more manual effort in terms of data pre-processing, data preparation, entity-relation extraction as well as feature extraction (See Figure 6). Furthermore, Knowledge Graph systems are more brittle in terms of the rules-based logic approach they use and are restricted by the knowledge of the designer. When applied to specific tasks in a single domain, they are easier to construct. However, if a manufacturing task required many-domain Knowledge to complete, then constructing a Knowledge Graph would be challenging especially in steps ② and ③ of Figure 6. This is especially true when context of an instruction is taken into account. Embedding context into Knowledge Graph and Natural Language computational models become very complex as the domains increase. As the number of domains required to complete a task increases, LLMs become the best option to use (see Figures 16 and 18). This is because their self-attention mechanisms gives them the capability to focus on the appropriate information from multiple domains. Nevertheless, the results in Figure 11 and Table 1 show that Pre-trained Large Models require large compute resources to host them as well as take longer when making an inference. This would make them difficult to be adopted by cash-strapped small businesses as well as increase the response time of collaborative robots respectively. As a result, it is beneficial to make use of lower compute models in these scenarios. However, this would require more manual effort in building such systems in order to ensure that the models generate the required information as needed by the domain. For example, in the Remanufacturing use case, the GPT models provided information on legislation that would need to be followed to carry out the task. This was an output that we did not foresee or program into the GPT models. Embedding this type of information or knowledge into NLP or WordNet models would be challenging to do.

## 6.2 Which Manufacturing Sector or Task stands to benefit most from applying Language Models (RQ3)?

| Sector | Machinery | Automotive | Electronics | Chemical | Plastics | Metal Fa... | Pharmac... | Aerospace | Wood | Furniture | Ceramics | Textile & ... | Food & B... | Sum |
|---|---|---|---|---|---|---|---|---|---|---|---|---|---|---|
| Machinery |  | ● | ● | ● | ● | ● | ● | ● | ● | ● | ● | ● | ● | 24 |
| Automotive | ● |  | ● | ● | ● | ● | ● | ● | ● | ● | ● | ● | ● | 24 |
| Electroni... | ● | ● |  | ● | ● | ● | ● | ● | ● | ● | ● | ● | ● | 23 |
| Chemical | ● | ● | ● |  | ● | ● | ● | ● | ● | ● | ● | ● | ● | 26 |
| Plastics ... | ● | ● | ● | ● |  | ● | ● | ● | ● | ● | ● | ● | ● | 26 |
| Metal Fa... | ● | ● | ● | ● | ● |  | ● | ● | ● | ● | ● | ● | ● | 19 |
| Pharmac... | ● | ● | ● | ● | ● | ● |  | ● | ● | ● | ● | ● | ● | 17 |
| Aerospace | ● | ● | ● | ● | ● | ● | ● |  | ● | ● | ● | ● | ● | 21 |
| Wood | ● | ● | ● | ● | ● | ● | ● | ● |  | ● | ● | ● | ● | 15 |
| Furniture | ● | ● | ● | ● | ● | ● | ● | ● | ● |  | ● | ● | ● | 16 |
| Ceramics | ● | ● | ● | ● | ● | ● | ● | ● | ● | ● |  | ● | ● | 13 |
| Textile & ... | ● | ● | ● | ● | ● | ● | ● | ● | ● | ● | ● |  | ● | 14 |
| Food & B... | ● | ● | ● | ● | ● | ● | ● | ● | ● | ● | ● | ● |  | 16 |

*Figure 17. A dependency matrix showing the dependency between the sectors of Machinery, Automotive, Electronics, Chemical, Plastics, Metal Fabrication, Pharmaceutical, Aerospace, Wood, Furniture, Ceramics, Textiles and Apparel, as well as Food and Beverages. In this matrix, a blue dot (which we gave a value of 3) means high dependency, a cyan dot (a value of two) means medium dependency while a red dot (with a value of one) means low dependency. Sectors like machinery have a medium to high dependency across all sectors while Textiles has a low to medium dependency across all sectors. Sectors like Electronics services a number of sectors such as Automotive, Chemicals and Plastics. Sectors like automotive with medium to high dependency across most sectors are likely to be more impacted by supply chain shocks. As a result, they are least resilient and need to be made more resilient through the use of remanufacturing techniques that are supported by Ontology or Large Language Models. With increasing customer personalisation, the automotive sector also serves as a prime candidate for introducing these models. In constructing this table, we made use of the 2024 data from the Bureau of Economic Analysis (BEA) [69]. The data from the BEA provides timely, relevant and accurate economic data of the United States. The data supports decisionmaking by businesses, policymakers and researchers. The sum column contains values of added across the other columns.*

Towards answering **RQ3** question, we made use of data from the Bureau of Economic Analysis to investigate how much domain or discipline knowledge is required to produce a product in various manufacturing sectors [69]. We also made use of data from the International Federation of Robotics to investigate the level of automation adoption in these sectors [70, 71]. The **Automotive sector**, for example, is highly automated, with robotics and automated assembly lines used extensively for tasks like welding and painting. Nevertheless, human operators are still needed for tasks requiring dexterity such as wiring and problem-solving. Furthermore, this sector is **highly dependent** on a complex supply chain of suppliers for various parts including electronics, plastics, machinery for the factories and metal fabrication to mention a few (See Figure 17). Also, the sector is faced with customers wanting ever more shorter lead times as well as personalised vehicles leading to more complexity in the manufacturing systems and supply chain. As a result, this sector is very prone to supply chain shocks .

On the other hand, a significant portion of the **wood and furniture sector** still relies heavily on manual labor due to the variability of wood and furniture designs across customers. In this sector, the human expertise required to create artifacts is focused within woodworking and furniture design. The sector also relies mostly on the machinery sector for power tools like milling as well as sawing and its dependency on other sectors is not as high as the automotive sector.

Furthermore, the use of automation is low due to the financial non-viability of developing automation for personalised and bespoke purposes in very low volumes. In the **consumer electronics sector**, automation is widely used to produce semi-conductors as well as parts for electronic components that go into devices across various industrial sectors. These components are produced en-massed with very little customisation. Furthermore, the supply chain that relies on these components would also be less resilient when compared with the wood or furniture sectors.

In the context of **resilience** in Industry 5.0, one way of *ensuring resiliency of supply chains is to make use of remanufacturing techniques in which local or regional factories specialising in circular economy are able to remanufacture an item to a near virgin condition*. These remanufactured items could then be used as buffers to weather supply chain shocks. Such a remanufacturing process would benefit from automation in disassembly, cleaning, and refurbishment processes for various parts. Nevertheless, human expertise would still be required to assess product condition, identify reusable components and ensure remanufactured products meet quality standards [73].

Towards this and to ensure resilient as well as sustainable business models, these local or regional factories would need to have access to a wide range of domain knowledge to be able to make use of raw materials from many recycled products in creating novel products. Consequently, robots in these factories would also need to have the ability to be *"cognitively flexible"* if they are to be used to produce remanufactured goods at scale. They would need to understand the various dependencies, *ethical and legislative considerations* to be considered during the remanufacture of goods. In achieving this, it is easier to make use of a computational model that has the capacity to capture the nuances across the domains and the links between them. As would be seen in the text generated by the computational models in Appendix C, GPT models have the capability to achieve this.

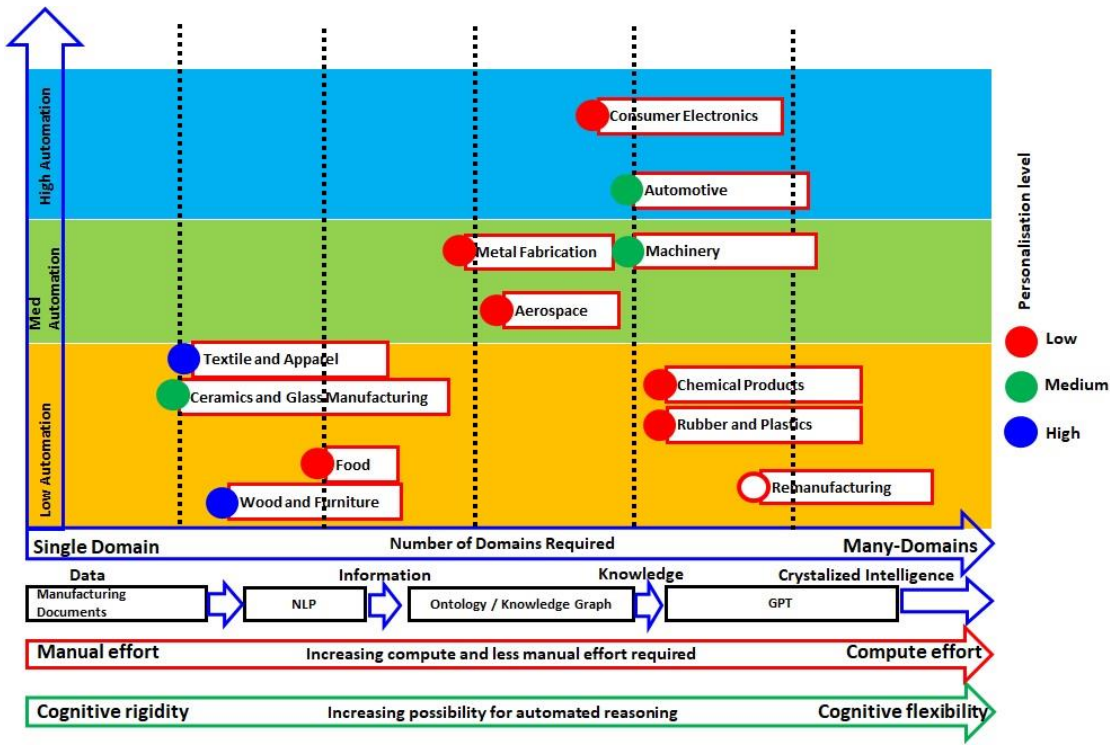

*Figure 18: A graph showing selected industrial sectors, usage of automation in each (for example, the automotive sector has a high level of automation compared to wood and furniture sector) and the number of domains (or sectors) from single to many domains required to produce a product **(Black horizontal arrow)**. The graph also shows how manual effort **(Red horizontal arrow)** reduces as more computational resources are used to train computational models. These computational models will increase in flexibility as their knowledge extends over multiple domains **(Green horizontal arrow)**. The methodologies used to produce computational models range from just manufacturing documents up to Generative Pre-Trained Transformer models (**Blue horizontal arrows** and **black boxes**). In deciding the level of automation in each selected sector, we made use of data from the International Federation of Robotics [67]. In cognitive psychology, two types of intelligence are defined in adult cognitive theories. These are fluid intelligence and crystallized intelligence. Fluid intelligence Crystallized intelligence involves the retrieval and application of previous knowledge and is used mostly for vocabulary knowledge, general information, and domain-specific skills. On the other hand, Fluid intelligence involves the rapid discovery and understanding of novel relationships as well as comprehension, reasoning and problem solving* [74].

This is because a Generative Pre-trained LLM has been trained on a variety of many-domain information and as a result, could potentially generate a variety of instructions for the construction of a variety of products from existing components. This opens up possibilities for the construction of novel products from disassembled components thereby enabling novel low volume and high variant resilient as well as sustainable manufacturing systems to be developed [75].

In such systems, a human interacting with and prompting a LLM intelligently could be used to generate bespoke instructions for each novel product type being manufactured. The human would then make use of their dexterity to solve manipulation problems that robots find difficult to do while collaborative robots would be used to conduct heavy, dirty and repeatable tasks. These tasks would change from novel product to product. This is unlike most current manufacturing systems in which a MES (Manufacturing Execution System) and a SCADA (Supervisory Control and Data Acquisition) system is programmed beforehand to hold every single command needed to assembly a discrete set of known goods.

However, in a LLM equipped factory, these instructions would be hosted in the architecture of the GPT potentially on a edge device thereby reducing the communication bandwidth as well as MES complexity required to achieve flexible manufacturing systems. Furthermore, achieving Ⓐ and Ⓑ in Figure 1 without an artificial cognitive system augmenting a human would lead to a highly complex manufacturing system that would rely heavily on human input and intervention [76, 77, 78]. These would lead to fatigue due to the need to keep changing actions for different product assemblies. It would also lead to higher error rates due to the limited mental bandwidth of humans. Overall, this would lead to inefficiency and a large reduction in productivity.

Furthermore, in the above vision of Cognitive Digital Job shops or Cognitive microfactories, humans will be able to use their cognition to intelligently prompt LLMs to create suggestions for assembling novel products from recycled materials. This would make such manufacturing systems human-centric thereby achieving another of the main goals of Industry 5.0 (See Figure 2). The ability to recycle/remanufacture products would also support supply chain resilience thereby addressing **RQ4**. The above possibility would complete the cycle shown in Figure 1 and begin another manufacturing evolution cycle that starts with Cognitive Digital job shops or Cognitive microfactories populated by cells of human-in-the-loop cognitive and reasoning robots.

### *6.3 Cognition, Reasoning and Intelligence in computational models*

In order to decide if a computational model is actually carrying out reasoning, it is important to define what reasoning is from a biological perspective. Cognition is the processing of information from perception all the way to action. Reasoning is the mental process of applying already known facts, rules, or experience to: **(i)** draw conclusions; **(ii)** make inferences based on information, evidence, or principles and **(iii)** to figure out something new or to evaluate a claim. This process could involve: **(i)** the construction of hypothesis based on current observations and then using known rules to make inferences **(Abductive)**; **(ii)** moving from specific observations to wider generalisations **(Inductive)**; or **(iii)** starting from general principles to make a decision on a specific matter **(Deductive)**. This is often conducted in real-time by a biological organism when presented with observations or tasks in the environment. The corresponding action of the organism leads to the observation of intelligence [79, 80].

When a task requires solving without prior knowledge on that task, then such intelligence is known as fluid intelligence while if solving the task makes heavy use of a prior known knowledge base, then it is known as crystallised intelligence. As a result, in demonstrating intelligence, there needs to be the process of real-time (flexible) reasoning that adapts to situations as well as the application (Crystallised) or non-application (Fluid) of a knowledge base [81].

In terms of Generative Pretrained Transformers, they rely heavily on an existing and very wide knowledge base upon which they have been trained. As a result, they fall in the crystallised intelligence spectrum. This wide knowledge base is then used to generate text that provides plausible responses to prompts given by humans. Depending on the use case and prompts given by humans, the GPTs can generate text that: **(i)** generalises from their knowledge base thereby simulating inductive reasoning; **(ii)** simulate deductive reasoning by making use of general knowledge to more specific use cases and **(iii)** forms hypotheses to best explain a series of observations thereby simulating Abductive reasoning. The key in all these, is the ability to do this on-the-fly and in real-time . Based on the above explanation, small-scale rule-based systems do not simulate reasoning because they make use of static rules that have been pre-programmed into them and hence cannot adapt to novel situations or changes in legislation [83].

### *6.4 Ethical and Legislative considerations*

Ethical and legislation considerations need to be taken into account during the production of goods. Different countries or economic regions have different legislation that need to be addressed if goods are to be exported to those countries. For example, the European Union has a different legislation to the United Kingdom and the United States on Electronics [84]. Furthermore, workplace ethical regulations are also different across countries [85]. This needs to be taken into consideration during the development of a computational model that supports the collaboration between humans and robots during the manufacture of a good. Embedding the regulations derived from legislation could be done using a rules based systems in single domains. However, as the number of dependent domains increase, it becomes increasingly difficult to capture this in such computational models. For example, in our experimental validation in remanufacturing, without explicitly mentioning safety or legislation to the GPT models, they were able to generate text that described the automotive regulations that need to be followed for the completion of the task. The WordNet or NLP computational models were not able to achieve this. Nevertheless, it is difficult to gain visibility or transparency of what legislative and ethical knowledge are embedded within a GPT model.

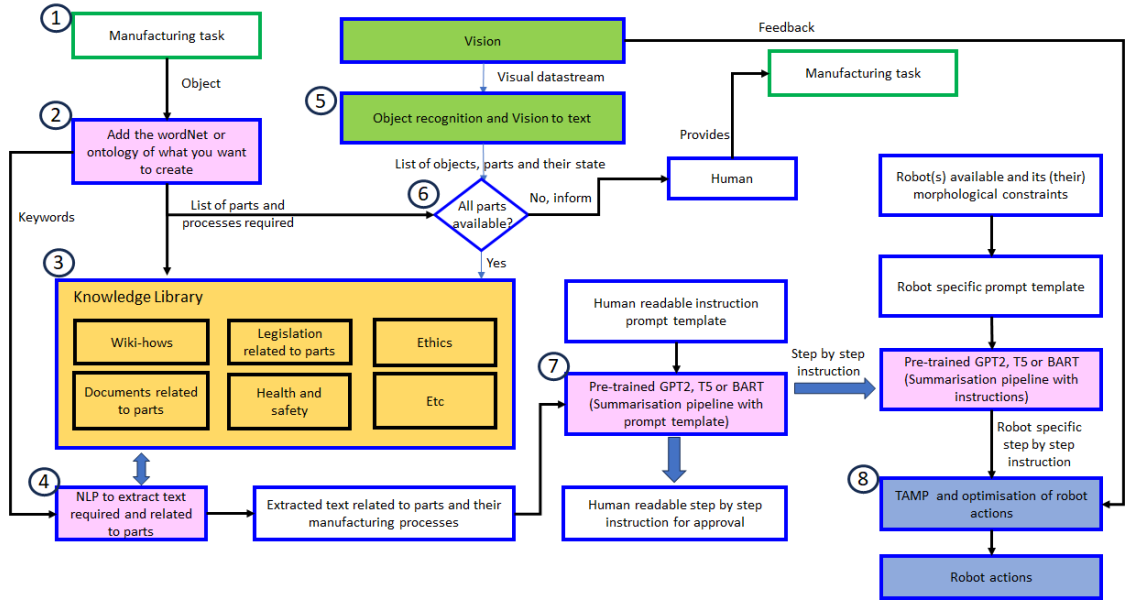

*Figure 19. Proposed flexible knowledge and Large Language Model Architecture for explanability and transparency across domains. The purple boxes are computatonal models. Together with Figure 18, this Architecture can be used as a decision guidance to decide where to put more manual effort where designing intelligent collaborative robots across sectors.*

### 6.5 Proposed Cognitive Architecture (RQ1) and Future work

Based on our findings above and inspired by Large Knowledge Language Models [86, 87], we propose the architecture shown in Figure 19. The architecture enables a human to define a manufacturing task ① as well as use the vast availability of Knowledge Graphs in literature to add the ontology of the task ② to the architecture. Furthermore, ③ enables users to add different types of supporting documents into a knowledge Library for use by the agent. These documents could include the readily available instructions on the internet in wikihows, documentations on legislation and ethics and so on. We propose the use of NLP techniques in ④ for the purposes of extracting relevant information from the Knowledge Library. Depending on the number of documents, the extracted textual data could be very large. As a result, a summarisation GPT in ⑦ is proposed. In preliminary experiments, we discovered that a text containing 63 words would take 11.33, 11.36 and 3.24 seconds for GPT2, BART (Bidirectional and Auto-Regressive Transformers) and T2 (Text-To-Text Transfer Transformer) architectures respectively. Since Pre-trained Transformers have the longest inference time and hence are the bottleneck, a T2 architecture could be used thereby making it more plausible to use on a robot.

The output of the chosen transformer will be guided by prompt templates that take human readability and robot morphology into account. In ⑧. Instructions will be used by a task and motion planning (TAMP) and optimisation pipeline to generate robot actions. Before the robot carries out the manufacturing task, it will use visual object recognition ⑤ to check if it has all the parts required to complete a task and feedback back to the human ⑥. The benefit of this architecture is that it provides visibility on how much manual or computation resources to devote to the modules in the architecture. This will depend on the complexity of the task as well as the domain in which the robot is operating (See Figure 17). It also provides transparency as to what knowledge the robot is using as informed by the documents contained in its Knowledge Library.

Furthermore, with the architecture, it is possible to configure a computational solution based on the level of computational resources available including whether to use T5 which take up 0.24 Gigabytes or BART which takes up 1.63 Gigabytes in storage. This architecture will be tested on various use cases in future work.

## Conclusion

In this work, we have discussed how Natural Language Processing techniques, Ontologies, Knowledge Graphs and most recently, Language Models have been integrated with industrial robotics. While these technology classes have different sub-classes (e.g GPT has many variants: Deepseek, Llama, Mistral etc), we recognise that the sub-classes of each class have commonalities in features, characteristics and techniques of training or building.

Using experimental validation, we have shown that despite the promise of Large Language Models, techniques such as Natural Language Processing and Ontologies can still be used in manufacturing with great benefits. We have also shown that industrial sectors that require the knowledge from many domains to manufacture a product stand to benefit the most from Large Language Models. We also signpost some current practices on how Large Language Models have been integrated with industrial robots and use that to inform which structure future Large Knowledge Language Models for manufacturing robots should take. Our work also provides a foundation for clarity and removes confusion between Natural Language and Language computational models for the benefit of the manufacturing community. Our work provides standards that can be used to inform decision makers looking to enhance their existing industrial systems with the latest Knowledge Augmented Language Models.

## Acknowledgement

This work was supported by the Engineering and Physical Sciences Research Council under grants: A digital COgnitive architecture to achieve Rapid Task programming and flEXibility in manufacturing robots through human demonstrations (DIGI-CORTEX) (EP/W014688/2)

## Appendix A: Generated text with scores for Textiles domain. A more detailed version is in the supplementary data.

| Model | Generated Text | Sce |
|---|---|---|
| gpt2-Beam Search | The cotton t-shirts are made from 100% cotton, and are available in a variety of colors and sizes. They are designed to be worn in the summer or in winter, or as a summer accessory. | 0 |
| gpt2-Greedy Search | Cotton fabric has a high density of polyesters, which are used to make the fabric. Polyester fabric also has high densities of fibers, and is used in the production of the cotton. Fiber production is the process of making a fabric that is resistant to moisture and moisture loss. Fibers are the fibers that are woven into the garment, the fabrics that hold the clothing together, or the materials that make up the garments. In the case of fabric, fiber is a material that has been woven to form a garment that can be worn. It is also a fiber that contains a number of other properties, such as a strong, elastic, stretchy, flexible, breathable, durable, waterproof. | 1 |
| gpt2-Random Sampling | In order to improve efficiency, you can also use all the above methods on your fabric. This also takes away unnecessary labor on the sewing machine. You can sew fabric into several sizes. Use our fabric size guide to find the right fabric. You can also choose any type of fabric to fit you. Materials used: Cotton Wool, Silky Silicone, Polyester Silk, Vinyl Silk, Natural Cotton, Lace (from our cotton collection), Alum-Cotton, Alum-Cotton, Fiber (and fabric colors listed below from our sources), or natural cotton from our plant To sew fabric you'll need to use an elastic or elastic lining, you'll need a sewing needle and a needle thread to sew it up. Next, make sure you get comfortable with your sewing machine. For most fabrics we use our nylon, we prefer nylon with a little. | 1 |
| gpt2-medium Beam Search | Cotton is the most versatile fabric in the world. It can be woven, woven with, or woven without. Cotton is also a very durable fabric, and can last for thousands of years. In fact, cotton is so durable that it is used to make clothing in many countries around the globe, | 0 |
| | including the U.S., Canada, Australia, New Zealand, Japan, South Korea, China, India, Indonesia, Malaysia, Thailand, Vietnam. | |
| gpt2-medium Greedy Search | The cotton shirt is made from cotton, cotton yarn, and cotton thread. The cotton is spun into yarn and woven into fabric. Cotton is a very durable material, but it is also very flammable. It is very difficult to burn cotton. If you are using cotton to make a shirt, you will need to use a flame retardant fabric, such as polyester, polypropylene, or polyethylene. Polyester is the most common flame-retardant material. You can also use polyvinyl chloride (PVC) or other flame resistant materials. | 1 |
| gpt2-medium Random Sampling | Construct a cotton T-shirt, starting from fiber production to fabric to finished garment. 4. Fabrication in Cotton-Making: 4.1 Making cotton cloth. 4.2 Fabricing cotton yarn or yarns for fabric manufacturing. 4.3 Fabrication of cotton rope or string. 5.1 Fabrication of clothing for wearing. 5.2 Fabrication of jewelry. 6.1 Fabrication of footwear. 7.1 Fabrication of furniture. 8.1 Fabrication of other textile material. 10.1 Fabrication of machinery and apparatus. 10.2 Fabrication of equipment. 11.1 Fabrication of other materials of construction and construction services. 12.1 Making clothes 13.1 Fabrication of clothing and footwear 14.1 Fabrication of garments. | 0.6 |
| gpt2-large Beam Search | Cotton is the most widely used fiber in the world. It is used for clothing, textiles, and other products. Cotton is also used as a feedstock for the production of plastics, pharmaceuticals, rubber, paper and many other items. In the United States, cotton is grown in more than 100 million acres of land, making it the second largest agricultural crop after corn. The U.S. Department of Agriculture (USDA) estimates that the cotton crop contributes $1.5 billion to the U.S. economy annually. | 0 |
| gpt2-large Greedy Search | The cotton is grown in the United States, and the cotton used in T shirts is imported from China. The cotton fiber is processed in China, where it is then shipped to the U.S. for processing. This process is called "fiber-to-fibers" (F2F).<br><br>"The F2Fs are then sent to a plant in Texas, which is where the finished T shirt is made. In the process, the F1F fiber from the T is converted into F3F fibers, then into the fiber-reinforced polyester (FRP) that is used to make the shirts. F4F is the raw material used for the final product."<br><br>In the past, cotton was grown on a large scale in India, but the country has been struggling with a drought. India is now looking to export its cotton to other countries. | 0 |
| gpt2-large Random | You can make your own.<br><br>You can find many ways to wear a cotton T-shirt. Some are simple and | 0 |

| | Sampling | some require skills, some cost money, some make you look better as a person, some are hard and require a lot of work, and some are difficult. | |
|---|---|---|---|
| | | You can make your own cotton T-shirt. It is not easy, but it is possible. | |
| | | If you want to learn how to create your own cotton T-shirt, you can do so. | |
| | | Why You Should Make A Cotton T-Shirt. | |
| | gpt2-xl Beam Search | Design and develop a new type of cotton fabric that can be used in a wide range of applications, including clothing, textiles, and other products. | 0 |
| | | Construct a cotton T-shirt, starting from fiber production to fabric to finished garment. | |
| | | Design and develop a new type of cotton fabric that can be used in a wide range of applications, including clothing, textiles, apparel, and other products. | |
| | gpt2-xl Greedy Search | Create a custom T shirt, from design to manufacturing to delivery. | 0 |
| | | We are a small team of designers and engineers who are passionate about creating great products. We are based in the UK and have a strong focus on customer service and quality. | |
| | gpt2-xl Random Sampling | Sew a shirt and a pair of jeans. Cut fabric and hang it. Sew another pair of jeans. Sew a pair of pants. Sew a shirt. | 0.6 |
| | | The above example is not an exhaustive list but it covers most of the basics. In the next two sections, we will look at the finer points, including creating and cutting fabric, fitting the shirt and jeans, sewing a shirt and pants, and more. | |
| | Word Net | Put together out of artificial or natural components or parts; He manufactured a popular cereal, a fabric woven from cotton fibers, and more. | 0 |
| | NLP-model | ['Cesses , mainly dyeing of fabrics , manufacture of synthetic fibers and spinning , directly affect the environment .', 'The fibers were fabricated by using hydrophobic Kevlar aerogel fibers (H-KAFs).'] | 0.3 |

## Appendix B: Generated text with scores for Electronics Domain. A more detailed version is in the supplementary data.

| Model | Generated Text | Sce |
|---|---|---|
| gpt2-Beam Search | In the early 1990s, IBM introduced the IBM PC, which was the first computer to run a graphical user interface (GUI) on a computer. The PC was designed to be portable and easy to use. It was also designed for use with a variety of other computers, including the Macintosh, Macintosh II, and Macintosh III. IBM also introduced a number of new computer operating systems, such as Windows, Mac OS X, Windows NT, Linux, FreeBSD, Solaris, NetBSD, OS/2, OpenBSD and OpenSolaris. In addition, the PC also had a built-in keyboard and mouse, as well as a touch-screen display. This made it possible for IBM to develop a wide range of products, from the personal computer, to the business computer and even the medical device. However, it was not until the mid-1990s that IBM began to make a significant investment in the development of its own operating. | 0.5 |
| gpt2-Greedy Search | The first computer to be built was the IBM PC-9800, which was built in the late 1960s. It was a small, but powerful computer, with a single floppy disk and a hard drive. The IBM computer was designed to run on a computer with an operating system called the Macintosh. In the early 1970s, IBM began to build computers with the Mac OS X operating systems. IBM's computer architecture was based on the Unix operating System, and it was called IBM-DOS. This was followed by the Windows operating Systems, the Linux operating operating set, Windows NT, Unix, Linux, FreeBSD, Solaris, MacOS X, OSX, NetBSD, OpenBSD and the Open Source operating sets. These operating environments were called Unix-like operating platforms. They were designed for use with computers running on computers that were not running the operating environment. For example, a Windows computer could run a DOS-. | 0.5 |
| gpt2-Random Sampling | Assembly a mainframe from electronic disks and keyboards to a full-fledged PC. While the company continues to use that hardware in conjunction with an alternative computer, it hopes to use it for other types of work.<br><br>It was unclear why it wanted to do this to the computer, but I am pretty sure that all the money and the time would have been better spent elsewhere, I'll admit.<br><br>I suppose it would be nice if IBM made a desktop PC with a desktop processor instead of a desktop-sized CPU. But that's not what it does, it just has a larger, cheaper PC.<br><br>If this sounds like a lot, that is because it really is. The IBM-made computer.<br><br>Assembly a mainframe from electronic disks and keyboards to a 3D printer. He also used the same software to build a series of other projects, the same process, which he said was inspired by Google's | 0.5 |
| | Mechanical Turk program. | |
| gpt2-medium Beam Search | A computer that can be used to create and edit computer programs. A computer program is a set of instructions that are executed by a computer to perform a specific task. For example, a program may be a text file that contains instructions for creating a new computer, or it may contain instructions to install a software program on a particular computer. The computer may also be an operating system, such as Windows, that is installed on the computer's hard drive.<br><br>In addition, computers may include software programs that perform specific tasks. Examples of computer software include operating systems, operating programs, and applications. | 0.8 |
| gpt2-medium Greedy Search | The first computer to use a microprocessor was the IBM PC, which was released in 1984. The first personal computer was a Commodore 64, released by Commodore in 1986. In 1987, the first home computer, a Sinclair ZX Spectrum, was launched. By 1988, there were more than 100,000 personal computers in use. Today, more personal computing devices are available than any other type of computer. | 0 |
| gpt2-medium Random Sampling | Assembly a mainframe from electronic disks and keyboards. This included a number of different models in different color schemes, but mostly in black. They were also manufactured in huge numbers as many as 4m per model. Their range would grow into the first few years of the 20th century, eventually reaching a peak of several thousands.<br><br>By 1980, IBM had expanded into the computer industry, now numbering about 50 companies including HP, EMC, NEC, and many more. At the same time, the technology was becoming increasingly powerful so the big companies began to consider computerized workstations. | 0 |
| gpt2-large Beam Search | The first computer to run on a floppy disk was the IBM PC, which was released in 1981. The PC was designed by a team of engineers at Bell Labs, and it was based on the DEC PDP-1, a single-board computer that had been designed in the early 1960s. It was a very small computer, with a 1.5-inch floppy drive and a 16-bit processor. In addition, the PC had a built-in keyboard and mouse, as well as a printer and printer port. This computer was so small that it could fit in a shoebox. | 0.5 |
| gpt2-large Greedy Search | The mainframes were used to create the first computer programs. The first program was a program called "The Game of Life." It was written by a man named John von Neumann. It is a game that is played by the computer. In the game, the player is given a number and has to find a way to get to the next number. If the number is not | 0 |
| | found, then the program stops. This program is called the "game of life." | |
| gpt2-large Random Sampling | A computer that displays a three-dimensional image of a person's face and body. The image can then be converted to color, a third dimension that is necessary for human vision. A computer that simulates a "third eye" to interpret information like face. A computer that interprets text, like computer programs. A computer that creates mathematical equations from binary binary numbers. A computer that processes data from a magnetic disk. A computer that creates data images from light pulses. In this example, the word "computer" is referring to the computer's mainframe. "A three-dimensional computer" is an alternate term for the same thing, but in this case it refers to the computer's graphical display. | 0 |
| gpt2-xl Beam Search | The mainframes were designed to be modular, so that they could be reconfigured to meet the needs of different users. The mainframes were also designed with a modular design in mind. This meant that the main frame could easily be replaced by a new one if it became damaged. In addition, the modularity of the system allowed for the creation of a wide variety of applications. For example, a user could create a program that would allow the user to control a computer from a remote location. Another application could allow a person to use the computer to create and edit a document. A third application would be used to manage a group of computers, such as a school or a company. These applications could then be connected together to form a larger system. | 0.5 |
| gpt2-xl Greedy Search | The mainframes were built by the IBM Corporation, and the mainboard was built at the Bell Telephone Laboratories. The mainboards were designed to be compatible with the computers that were being built in the early 1960s. This meant that the computer could be connected to the keyboard and used as a keyboard. It was also designed so that it could run a program written in a language called Basic. Basic was a programming language that was designed for use on the Apple II computer. In order to run the program, the user would have to enter commands into the machine. | 0.5 |
| gpt2-xl Random Sampling | Ans: In addition to creating an operating system, you can add a compiler or assembler to the system. The standard Unix tools – gcc – are the right tool for writing the first compiler (the assembler is a compiler by itself), but a compiler for a particular language can be written in a different language. Assembly a mainframe from electronic disks and keyboards or other devices, such as a game pad, and create a mainframe from any digital material. Cisco will be joining industry leader IBM and technology pioneer Ericsson as partners in the program. Other firms are expected to join the first round of companies when they are announced. | 0.5 |

| | The program will be focused on developing new hardware, software and systems architecture. The goal is to create products that are cheaper, faster and easier to use than existing versions. | |
|---|---|---|
| Word Net | Assembly a a large digital computer [a machine for performing calculations automatically] serving [be used by; as of a utility] 100-400 users [a person who makes use of a thing; someone who uses or employs something] and occupying [march aggressively into another's territory by military force for the purposes of conquest and occupation] a special air-conditioned room [the people who are present in a room] from electronic sound [(phonetics) an individual sound unit of speech without concern as. | 0.8 |
| NLP-model | ['Further information can migrate from and to the user's microcomputer and host mainframe.', 'It was strongly emphasized that CAPP systems should be built in such a PCs, workstations, or mainframes.', 'In addition, the software itself was often highly specialized for the mainframe operating systems it served and the cost of software was prohibitive to the normal buyer.', 'As relevant to the market of mainframe computers.', 'Resource costs will be assigned to an activity proportional to the value the activity has on the resource "mainframe" is "CPU seconds".', 'Each activity use and the costs of the mainframe will be divided according to the activity's share of the total CPU seconds used.'] | 0.4 |

## Appendix C: Generated text with scores for Remanufacturing for the Automotive Domain. A more detailed version is in the supplementary data.

| Model | Generated Text | Sce |
|---|---|---|
| gpt2-Beam Search | The Secretary of Transportation shall establish a program to provide for the conversion of an existing electric motor vehicle to a hybrid vehicle by the Secretary. The program shall include the following elements: (A) A plan to convert an old electric motorcycle into a new electric bicycle. (B) An estimate of the cost of converting an older electric bike into the new bicycle, including the costs of replacing the old bicycle with a replacement bicycle and any other costs that may be incurred to replace the existing bicycle or replace a bicycle that has been damaged or destroyed by an accident…. | 0.5 |
| gpt2-Greedy Search | The vehicle must be equipped with a power steering system that is capable of handling up to 100,000 pounds of torque. The vehicle's power output must not exceed 100 horsepower. A vehicle equipped to drive an automatic transmission must have a maximum power rating of 100 hp. An electric motor must also be capable to handle up to 100 pounds. | 1 |
| gpt2-Random Sampling | (See section 3.10, "Electric vehicle fuel tax on all motor vehicles" in the final rulemaking note) 5. An engine vehicle not owned or leased by the user may not be converted into an electric vehicle while parked within the limits of this regulation unless it has met all requirements set forth in section 2 of the Environmental Protection Act. A vehicle of this class may not be converted into an electric vehicle while parked inside the limits of this regulation. 6. If an electric vehicle is owned or leased under this chapter and any restrictions imposed under this chapter do not apply to it, the operator of the vehicle must immediately turn the vehicle into an electric vehicle. 7. Whenever an electric vehicle that is registered with the Department is converted into an electric vehicle pursuant to paragraph 4 (a) of subdivision 1, subparagraph (B). | 0.5 |
| gpt2-medium Beam Search | Electric Vehicle Charging Station (EVDS) - A charging station for electric vehicles that can be used to charge a vehicle's battery. The EVDs are designed to be portable and easy to use. They can also be connected to a car's electrical system to provide charging for the car. Electronic Vehicle (EV) Charger - An electronic vehicle charger that provides charging to an EV's electric motor. It is a plug-in hybrid vehicle charging system that uses a battery pack to power the electric motors. This system is designed for use with electric cars and is not compatible with gasoline-powered vehicles. EV chargers are available for sale in the United States and other countries, including Canada, Europe, Japan, and South Korea. In addition, EV charging stations are being developed in China, India, Mexico, South Africa, the Middle East and Africa. | 0.5 |
| gpt2-medium Greedy Search | The vehicle must be capable of operating on a highway, and must meet all applicable federal, state, local, tribal, or international safety standards. The vehicle may not be used for commercial purposes. (Sec. 5) Requires the Secretary to establish a program to develop and implement a national electric mobility strategy. Requires each State to submit to the President a plan for the development and implementation of the strategy by the end of FY2018. (Sec, 523) Prohibits funds from being used to implement the National Electric Mobility Strategy until the Administrator of General Services submits to Congress a report on the implementation and effectiveness of that strategy and the results of any studies conducted by that agency. | 0.5 |
| gpt2-medium Random Sampling | Convert an internal combustion engine vehicle into an electric vehicle. These are all great tools, and they allow manufacturers to maximize efficiency. However, there are some key challenges for EVs in the US and Europe.<br><br>Here are some key questions about the EU and the US that I've fielded over the years: EU EV requirements On the US side of the Atlantic, the EIA and NREL published detailed vehicle performance specifications | 0.5 |
| | for various European EV models. These show what each model requires and what the specific engine size is. These specs are typically available with the model years. However, they don't contain the technical specifications of each EV in the US. | |
| gpt2-large Beam Search | A vehicle is an engine-powered vehicle that is designed to travel on a public highway. A vehicle may be used to transport passengers, goods, or both. The term "vehicle" includes a motor vehicle, trailer, semitrailer, truck, bus, motorcycle, moped, and any other vehicle designed or used for the transportation of persons or property on the public highways of this state. | 0.5 |
| gpt2-large Greedy Search | The conversion process is simple and can be performed in a matter of minutes. The conversion is performed by using a high-voltage battery pack, which is connected to the engine via a direct current (DC) power supply. This DC power is supplied to a DC motor, and the motor is driven by a motor controller. A DC electric motor can drive a vehicle up to 100 km (62 miles) on a single charge. In addition, the DC motors can also be used to power a generator, or to provide power to an inverter. | 1 |
| gpt2-large Random Sampling | Convert a heavy-duty truck into an electric vehicle, including any heavy-duty part that can be made into a generator, for example, a diesel engine.<br><br>He and colleague Robert Cialdini, director of energy use at the University of Michigan, said a future Tesla, a Tesla Model S, and a Chevy Bolt have been tested in Michigan but they won't be the only electric vehicles. | 0 |
| gpt2-xl Beam Search | Convert an internal combustion engine vehicle into an electric vehicle. (1) A person may not convert a vehicle that is a motor vehicle or a trailer into another vehicle unless the person complies with the following requirements: (a) The vehicle must be equipped with an electrical system that meets the requirements of 49 C.F.R. Part 383, Subpart E. (b) If the conversion is to be done by a person other than the manufacturer of the vehicle, the electric system must have a capacity of not less than 1,000 watts and a maximum output of at least 1.5 kilowatts, and (c) In the case of a conversion to a plug-in hybrid electric drive system, the battery pack must not be larger than 2.4 kWh. | 1 |
| gpt2-xl Greedy Search | The first step is to convert the engine into a generator. The generator is then connected to the battery pack. This is done by connecting the generator to a battery. A battery is a device that stores energy. It is made up of a positive and a negative electrode. When the positive electrode is charged, it releases electrons. These electrons are then stored in the negative battery electrode, which is connected with the motor. As the vehicle drives, the electrons flow through the electric | 1 |
| | motor, producing electricity. In this way, a vehicle can be powered by a combination of electricity and gasoline. | |
| gpt2-xl Random Sampling | The purpose of the aerodynamics is to minimize the forces on the car, by minimizing drag. The air is a compressible solid, so it can be compressed at high velocity to change shape. The change in shape is not enough to produce a vortex; a vortex is generated with high velocity when a solid or liquid is compressed, and the speed of the compressible solid is increased in the direction of the vortex. To learn more about our program or to apply please go to our Apply Now! page. | 0 |
| Word Net | Convert an internal a state [(chemistry) the three traditional states of matter are solids (fixed shape and volume) and liquids (fixed volume and shaped by the container) and gases (filling the container)] of violent disturbance [activity that is a malfunction, intrusion, or interruption] and excitement [the state of being emotionally aroused and worked up] a wheeled vehicle [any substance that facilitates the use of a drug or pigment or other material that is mixed with it]. | 0.5 |
| NLP-model | Comparative case study of internal combustion engine (ICE) and electric car. | 0 |